\documentstyle[12pt]{article}
\input mssymb
\begin{document}
{\noindent \Large\bf On the Obstructions to non-Cliffordian}\\
\vspace*{0.18cm}

{\noindent \Large\bf Pin Structures}\\
\vspace*{0.4cm}

{\noindent \bf Andrew Chamblin}\\
\vspace*{0.15cm}

{\noindent Department of Applied Mathematics and Theoretical Physics,
University of
Cambridge, Cambridge CB3 9EW, England}\\

\vspace*{0.6cm}
{\noindent \bf Abstract}\\

{We derive the topological obstructions to the existence of non-Cliffordian pin
structures on four-dimensional spacetimes. We apply these obstructions to the
study of
non-Cliffordian pin-Lorentz cobordism. We note that our method of derivation
applies
equally well in any dimension and in any signature, and we present a general
format for
calculating obstructions in these situations. Finally, we interpret the
breakdown of pin
structure and discuss the relevance of this to aspects of physics.}\\

\vspace*{0.6cm}
{\noindent \bf Contents}\\

{\noindent I.  ~Introduction}\\
\\
{\noindent II. ~Discussion of sheaf theory}\\
\\
{\noindent III. ~Derivation of the obstructions to non-Cliffordian pin
structures}\\
\\
{\noindent IV. ~Applications of the obstructions to pin-Lorentz cobordism}\\
\\
{\noindent V. ~Interpreting the breakdown of pin structure}\\
\\
{\noindent VI. ~Format for solving the general problem}\\
\\
{\noindent VII. ~Conclusion}\\
\\
{\noindent Acknowledgements}\\
\\
{\noindent References}\\

\vspace*{0.6cm}
{\noindent \bf I. ~Introduction}\\

{Suppose we are given a manifold, $M$, with tangent bundle ${\tau}_{M}$ which
can be
reduced to a bundle with structure group `$O$', say.  Then one of the first
things we
might notice is that we generically have ${\pi}_{1}(O) ~{\simeq}~ G
{}~{\not\simeq}~
{\{}1{\}}$. What this means is that at a point $p ~{\in}~ M$ there exist paths
$O_{1},
O_{2} ~{\in}~ O$, which might act on the fibre ${\tau}_{M}|_{p}$ `equivalently'
(in
the sense that, for $x ~{\in}~ {\tau}_{M}|_{p}, O_{1}(x) = O_{2}(x)$), but with
the
property that $O_{1}$ and $O_{2}$ (viewed as curves in $O$) are not homotopic,
i.e.,
cannot be continuously deformed into each other. This might disturb us, and so
we may
be inclined to represent the information contained in the tangent bundle in a
{\it simply connected} manner. What this amounts to locally (in a
neighbourhood about $p$) is finding some bundle ${\varsigma}_{M}$, with
structure group
${\bar O}$ given by the exact sequence $1 ~{\longrightarrow}~ {\pi}_{1}(O)
{}~{\longrightarrow}~ {\bar O} ~{\longrightarrow}~ O ~{\longrightarrow}~ 1$.
Then
locally the bundle ${\varsigma}_{M}$ `encodes' all of the information that was
contained in ${\tau}_{M}$. However, we may not be able to find such a bundle
{\it globally}, i.e., there are topological obstructions to globally
`re-representing' the information of ${\tau}_{M}$ in a simply connected way.

In this paper, we are going to concentrate on {\it spacetimes}, $M$, which are
not necessarily orientable. What this means is that the tangent bundle,
${\tau}_{M}$,
can at most be reduced to an $O(p, q)$ bundle. When the metric, $g_{ab}$, has
signature $(- + + +)$ then the structure group will be $O(3, 1)$. When the
metric has
signature $(+ - - -)$ then the structure group will be $O(1, 3)$ (actually,
$O(3, 1)
{}~{\simeq}~ O(1, 3)$, but as we shall see it is necessary to keep the
distinction when
we pass to the double covers). Since ${\pi}_{1}(O_{0}(3, 1) ~{\simeq}~
{\pi}_{1}(O_{0}(1, 3)) ~{\simeq}~ {\Bbb Z_{2}}$, we are interested in finding
all groups
which are double covers of $O(3, 1)$ and $O(1, 3)$. However, there are
{\it eight} distinct such double covers [2] of $O(p, q)$! Following Dabrowski,
we will write these covers as}
\[
{h^{a, b, c}:~ {\mbox{Pin}}^{a, b, c}(p, q) ~{\longrightarrow}~ O(p, q)}
\]
{with $a, b, c ~{\in}~ {\{}+, -{\}}$. The signs of $a, b$, and $c$ can be
interpreted
in the following way:

Recall, first, that $O(p, q)$ is not path connected; there are four components,
given
by the identity connected component, $O_{0}(p, q)$, and the three components
corresponding to parity reversal $P$, time reversal $T$, and the combination of
these
two, $PT$ (i.e., $O(p, q)$ decomposes into a semidirect product{\footnote{That
is,
$O(p, q)$ is the disjoint union $O(p, q)$ $=$ $(O_{0}(p, q))$ $~{\cup}~$
$P(O_{0}(p,
q))$ $~{\cup}~$ $T(O_{0}(p, q))$ $~{\cup}~$ $PT(O_{0}(p, q))$, and the four
element
group ${\{}1, P, T, PT{\}}$ is isomorphic to ${\Bbb Z_{2}} ~{\times}~ {\Bbb
Z_{2}}$.}}, $O(p, q) ~{\simeq}~ O_{0}(p, q) ~{\odot}~ ({\Bbb Z_{2}} ~{\times}~
{\Bbb Z_{2}})$). The signs of $a, b$, and $c$ then correspond to the signs of
the
squares of the elements in ${\mbox{Pin}}^{a, b, c}(p, q)$ which cover space
reflection,
$R_{S}$, time reversal, $R_{T}$ and a combination of the two respectively.
(Recall
that parity $P$ is written $P = R_{x}R_{y}R_{z}$, the product of reflections
about the
three spacelike axes).

With this in mind we can, following Dabrowski [2], write out the explicit form
of the
groups ${\mbox{Pin}}^{a, b, c}(p, q)$; they are given by the semidirect
product}
\[
{{\mbox{Pin}}^{a, b, c}(p, q) ~{\simeq}~ {\frac{({\mbox{Spin}}_{0}(p, q)
{}~{\odot}~
C^{a, b, c})}{{\Bbb Z_{2}}}}}
\]
{where the $C^{a, b, c}$ are the four double coverings of ${\Bbb Z_{2}}
{}~{\times}~
{\Bbb Z_{2}}$; i.e., $C^{a, b, c}$ are the groups ${\Bbb Z_{2}} ~{\times}~
{\Bbb
Z_{2}} ~{\times}~ {\Bbb Z_{2}}$ (when $a = b = c = +$), $D_{4}$ (dihedral
group, when
there are two plusses and one minus in the triple $a, b, c$), ${\Bbb Z_{2}}
{}~{\times}~
{\Bbb Z_{4}}$ (when there are two minuses and one plus in $a, b, c$), and
$Q_{4}$
(quaternions, when $a = b = c = -$). Interestingly, the only groups which can
be
obtained from the Clifford algebras $Cl(p, q)$ (in the usual way) are}
\[
{{\mbox{Pin}}^{+, -, +}(p, q) ~{\simeq}~ {\frac{({\mbox{Spin}}_{0}(p, q)
{}~{\odot}~
D_{4})}{{\Bbb Z_{2}}}} \hspace*{2.2cm} ~~~~~~~~~{\mbox{and}}}
\]
\[
{{\mbox{Pin}}^{-, +, +}(p, q) ~{\simeq}~ {\frac{({\mbox{Spin}}_{0}(q, p)
{}~{\odot}~
D_{4})}{{\Bbb Z_{2}}}} .}
\]
{These pin groups are therefore called `Cliffordian', and the obstruction
theory for
{\it Cliffordian} pin structure was worked out by Karoubi [3], see also [1].
We are concerned with the obstruction theory for the non-Cliffordian pin
groups. To
see how to approach this problem, let us first review the structures involved.

Recall, first of all, that $O(p, q)$ decomposes as a semidirect product $O(p,
q)$
$ ~{\simeq}~ O_{0}(p, q) ~{\odot}~ ({\Bbb Z_{2}} ~{\times}~ {\Bbb Z_{2}})$.
Likewise,
the pin groups decompose into semidirect products via ${\mbox{Pin}}^{a, b,
c}(p, q)
{}~{\simeq}~ {\frac{({\mbox{Spin}}_{0}(p, q) ~{\odot}~ C^{a, b, c})}{{\Bbb
Z_{2}}}}$,
where ${\mbox{Spin}}_{0}(p, q)$ is the $2 - 1$ cover of $O_{0}(p, q) ~{\simeq}~
SO_{0}(p, q)$ and $C^{a, b, c}$ are $2 - 1$ covers of ${\Bbb Z_{2}} ~{\times}~
{\Bbb Z_{2}}$. These semidirect products are naturally associated with the
homomorphisms}
\[ \left\{ \begin{array}{l}
h_{1}:~ C^{a, b, c} ~{\longrightarrow}~ {\mbox{Aut}}({\mbox{Spin}}_{0}(p, q))\\
h_{2}:~ {\Bbb Z_{2}} ~{\times}~ {\Bbb Z_{2}} ~{\longrightarrow}~
{\mbox{Aut}}(SO_{0}(p, q))
\end{array}
\right. \]
{i.e., for example, if $e_{1}$ represents time reflection in ${\mbox{Pin}}^{a,
b,
c}(p, q)$, then $h(e_{1})$ is equal to the map (automorphism) on
${\mbox{Spin}}_{0}(p,
q)$ given by conjugation:}
\[
{{\mbox{Spin}}_{0}(p, q) ~{\ni}~ v_{1} v_{2} ... v_{k} ~{\longrightarrow}~
e_{1} v_{1}
v_{2} ... v_{k} e_{1}^{-1}}
\]
{and similarly for $h_{2}$. In other words, if $({\varsigma}_{1}, c_{1})
{}~{\in}~
{\mbox{Spin}}_{0}(p, q) ~{\odot}~ C^{a, b, c}$ and $({\varsigma}_{2}, c_{2})
{}~{\in}~
{\mbox{Spin}}_{0}(p, q) ~{\odot}~ C^{a, b, c}$, then multiplication of the two
elements of the semidirect product is given by}
\[
{({\varsigma}_{1}, c_{1})({\varsigma}_{2}, c_{2}) = ({\varsigma}_{1} c_{1}
{\varsigma}_{2} c_{1}^{-1}, c_{1} c_{2})}
\]
{and so on.

What this means [4] is that we obtain exact sequences:}
\[ \left\{ \begin{array}{cccccccccc}
1 &{\longrightarrow} &{\mbox{Pin}}_{0}^{a, b, c}(p, q) &{\longrightarrow}
&{\mbox{Pin}}^{a, b, c}(p, q) &{\longrightarrow} &C^{a, b, c}
&{\longrightarrow}  &1
& \\
 & & & & & & & & &\hspace*{0.4cm} ~~\hfill (1)\\
1 &{\longrightarrow} &O_{0}(p, q) &{\longrightarrow} &O(p, q)
&{\longrightarrow}
&{\Bbb Z_{2}} ~{\times}~ {\Bbb Z_{2}} &{\longrightarrow} &1 &
\end{array}
\right. \]

{Furthermore, because the elements of the top sequence are $2 - 1$ covers of
elements
of the bottom sequence, we see that we must have the following (commutative)
diagram:}
\[
\begin{array}{cccccccccc}
1 &{\longrightarrow} &{\mbox{Pin}}_{0}^{a, b, c}(p, q) &{\longrightarrow}
&{\mbox{Pin}}^{a, b, c}(p, q) &{\longrightarrow} &C^{a, b, c}
&{\longrightarrow}
&1 & \\
 & &{\downarrow} & &{\downarrow} & &{\downarrow} & & &\hspace*{0.45cm}
{}~~~\hfill (2) \\
1 &{\longrightarrow} &O_{0}(p, q) &{\longrightarrow} &O(p, q)
&{\longrightarrow}
&{\Bbb Z_{2}} ~{\times}~ {\Bbb Z_{2}} &{\longrightarrow} &1 &
\end{array}
\]

{Thus, diagram (2) `fixes' the structure of ${\mbox{Pin}}^{a, b, c}(p, q)$
given
$C^{a, b, c}$.

Including the short exact sequences which `express' the fact that
${\mbox{Pin}}^{a, b,
c}(p,$  $q)$ and ${\mbox{Pin}}_{0}^{a, b, c}(p, q)$ are $2 - 1$ covers of $O(p,
q)$ and
$O_{0}(p, q)$, we obtain the commutative diagram}
\[
\begin{array}{cccccccccc}
 & &| & &| & &| & & & \\
 & &{\downarrow} & &{\downarrow} & &{\downarrow} & & & \\
 & &{\Bbb Z_{2}} & &{\Bbb Z_{2}} & &{\Bbb Z_{2}} & & & \\
 & &{\downarrow} & &{\downarrow} & &{\downarrow} & & & \\
1 &{\longrightarrow} &{\mbox{Pin}}_{0}^{a, b, c}(p, q) &{\longrightarrow}
&{\mbox{Pin}}^{a, b, c}(p, q) &{\longrightarrow} &C^{a, b, c}
&{\longrightarrow} &1 &
\\
 & &{\downarrow} & &{\downarrow} & &{\downarrow} & & &\hspace*{0.45cm} ~~\hfill
(3) \\
1 &{\longrightarrow} &O_{0}(p, q) &{\longrightarrow} &O(p, q)
&{\longrightarrow}
&{\Bbb Z_{2}} ~{\times}~ {\Bbb Z_{2}} &{\longrightarrow} &1 & \\
 & &{\downarrow} & &{\downarrow} & &{\downarrow} & & & \\
 & &| & &| & &| & & &
\end{array}
\]

{At first glance, the above diagram looks innocuous. However, as we shall see,
when we
view the algebraic structures in the diagram as {\it sheaves}, we will obtain
a commutative diagram of sheaves, from which we will obtain a commutative
diagram of
sheaf cohomology groups, with which we will be able to derive our obstructions.
Before
we do this, however, it is useful to review sheaf cohomology.}\\
\vspace*{0.6cm}

{\noindent \bf II. ~Discussion of sheaf{\footnote{This discussion is taken
primarily
from Wells [5], Chapter II.}} theory}\\

{`Sheaf theory' is, broadly speaking, a mathematical technology that allows us
to
connect information which is local with information which is global. A
{\it sheaf} is roughly something that tells us about localized information on
$M$. To pass to global information, we need sheaf cohomology.

To make this more precise, let $M$ be a topological space. Then a
{\it presheaf} $S$ over $M$ is an assignment of a set $S(U)$ to every
non-empty set $U ~{\subset}~ M$, such that for every pair of open sets $U_{1}
{}~{\subset}~ U ~{\subset}~ M$ we have {\it restriction homomorphisms}
$r_{U_{1}}^{U}:~ S(U) ~{\longrightarrow}~ S(U_{1})$ which satisfy}\\

{\noindent (a) $r_{U}^{U} =$ `identity map on $U$'}\\

{\noindent (b) For any open sets~ $U_{2} ~{\subset}~ U_{1} ~{\subset}~ U$,
$r_{U_{2}}^{U} = r_{U_{2}}^{U_{1}} ~{\bf {\circ}}~ r_{U_{1}}^{U}$.}\\
\\
{\noindent {\bf Definition}~~ Let ${\cal A}$ and ${\cal B}$ be presheaves
over $M$. Then we define a {\it morphism} of presheaves to be a set of mappings
$f_{U}:~ {\cal A}(U) ~{\longrightarrow}~ {\cal B}(U)$, for each open set $U
{}~{\subset}~ M$, such that the diagram}
\[
\begin{array}{ccc}
{\cal A}(U) &{\longrightarrow} &{\cal B}(U) \\
 & & \\
{\downarrow} r_{U_{1}}^{U} & &{\downarrow} r_{U_{1}}^{U} \\
 & & \\
{\cal A}(U_{1}) &{\longrightarrow} &{\cal B}(U_{1})
\end{array}
\]

{\noindent is commutative, where $U_{1} ~{\subset}~ U ~{\subset}~ M$, $U_{1}$
open. We
write such a morphism as $f:~ {\cal A} ~{\longrightarrow}~ {\cal B}$.

Let ${\{}U_{i}{\}}$ be any collection of open subsets of $M$ such that $U =
{\stackrel{\bigcup}{i}} U_{i}$. A presheaf ${\cal A}$ is a {\it sheaf} iff it
satisfies the following two `Sheaf Axioms':}\\
\\
{\noindent {\bf Axiom 1}~ {\it If $a, b ~{\in}~ {\cal A}(U)$ and ${\forall} ~i,
{}~r_{U_{i}}^{U}(a) = r_{U_{i}}^{U}(b)$, then $a = b$.}}\\
\\
{\noindent {\bf Axiom 2}~ {\it If for $a_{i} ~{\in}~ {\cal A}(U_{i})$ and
$U_{i} ~{\cap}~ U_{k} ~{\not=}~ {\emptyset}$ we have}}
\[
{r_{U_{i} ~{\cap}~ U_{k}}^{U_{i}}(a_{i}) = r_{U_{i} ~{\cap}~
U_{k}}^{U_{k}}(a_{k})}
\]
{{\it for any $i$, then there exists $a ~{\in}~ {\cal A}(U)$ such that
$r_{U_{i}}^{U}(a) =
a_{i}, {\forall} ~i$.}}\\
\\
{Intuitively, Axiom 1 says that sheaves encode their information locally,
whereas Axiom
2 says that we can `piece together' local information to get global
information.

A {\it mapping} of sheaves, ${\cal A} ~{\longrightarrow}~ {\cal B}$, is a
morphism of the underlying presheaves.

Now, there are many interesting examples of sheaves and their applications in
geometry
and mathematical physics, and we refer the reader to [5] and [6] for a thorough
treatment. For our purposes, we shall be concerned with {\it constant}
sheaves, i.e., sheaves which are simply the assignment $U ~{\longrightarrow}~
{\cal
G}$ of some group ${\cal G}$ to any connected open set $U ~{\subset}~ M$.

Consider, now, the structure ${\cal A}_{x}$ obtained from a sheaf, ${\cal A}$
via}
\[
{{\cal A}_{x} = {\lim_{{\rightharpoonup} \atop x {\in} U_{i}}} {\cal A}(U_{i})}
\]
{where `${\displaystyle {\lim_{{\rightharpoonup} \atop x {\in} U}}}$' refers to
the
direct limit of the restriction homomorphisms over nested neighbourhoods $U_{1}
{}~{\subset}~ U_{2} ~{\subset}~ ... ~{\subset}~ U_{i} ~{\subset}~ ...$ about
$x$. Then
${\cal A}_{x}$ is called the {\it stalk} of ${\cal A}$ at $x ~{\in}~ M$.

If ${\cal A}, {\cal B}$, and ${\cal C}$ are sheaves of groups on $M$, the
sequence of
morphisms}
\[
{{\cal A} ~{\stackrel{\mu}{\longrightarrow}}~ {\cal B}
{}~{\stackrel{\nu}{\longrightarrow}}~ {\cal C}}
\]
{is {\it exact} if the corresponding sequence on stalks}
\[
{{\cal A}_{x} ~{\stackrel{\mu_{x}}{\longrightarrow}}~ {\cal B}_{x}
{}~{\stackrel{\nu_{x}}{\longrightarrow}}~ {\cal C}_{x}}
\]
{is exact ${\forall} ~x ~{\in}~ M$. A short exact sequence is a sequence of
morphisms}
\[
{1 ~{\stackrel{f}{\longrightarrow}}~ {\cal A} ~{\stackrel{g}{\longrightarrow}}~
{\cal
B} ~{\stackrel{h}{\longrightarrow}}~ {\cal C} ~{\stackrel{j}{\longrightarrow}}~
1
\hspace*{5.75cm} ~~~~~~~~~~~~\hfill (4)}
\]
{with ${\mbox{Im}}(f) = {\mbox{ker}}(g), {\mbox{Im}}(g) = {\mbox{ker}}(h),
{\mbox{Im}}(h) = {\mbox{ker}}(j)$. {\it Sheaf cohomology} is, roughly
speaking, concerned with measuring `how exact' (4) is, i.e., to what extent
${\mbox{Im}}(h) ~{\not=}~ {\mbox{ker}}(j)$. We now can develop sheaf cohomology
theory
[6] from the `${\breve{\rm C}}$ech' point of view. The point now is that the
coefficients for the cohomology will be sections of the sheaf, $S$, in question
(i.e.,
sections are elements of $S(U)$). That is to say, we view (${\breve{\rm
C}}$ech)
$q$-cochains as maps $C^{q}:~ U_{0} ~{\cap}~ U_{1} ~{\cap}~ ... ~{\cap}~ U_{q}
{}~{\longrightarrow}~ S(U_{0} ~{\cap}~ U_{1} ~{\cap}~ ... ~{\cap}~ U_{q})$,
where
$U_{0}, U_{1}, ... U_{q}$ are $q ~+~ 1$ open sets in $M$ with non-empty
intersection.
We can define a coboundary operator, ${\delta}:~ C^{q} ~{\longrightarrow}~ C^{q
{}~+~
1}$, in the usual way and so in an appropriate limit [6] we get the {\it sheaf
cohomology groups of $M$ with coefficients in $S$:}}
\[
{H^{*}(M; S) .}
\]

{Now, since our sheaves are all going to be constant, this cohomology will in
fact
reduce to the usual cohomology.

We now state the main result which we will need to calculate the obstructions
in the
next section:}\\
\\
{\noindent {\bf Theorem}~[5] {\it ~Let $M$ be Hausdorff and
paracompact{\footnote{Recall that by Geroch [7] all spacetimes have these
properties.}}. Then}}\\

{\noindent {\it (a) For any sheaf ${\cal A}$ over $M$,}}
\[
{H^{0}(M; S) = {\Gamma}(M; S) = {\mbox{\it `sections of $S$ over $M$'}}}
\]

{\noindent {\it (b) For any sheaf morphism}}
\[
{h:~ {\cal A} ~{\longrightarrow}~ {\cal B}}
\]
{{\it there is, for $q ~{\geq}~ 0$, a group homomorphism}}
\[
{h^{q}:~ H^{q}(M; {\cal A}) ~{\longrightarrow}~ H^{q}(M; {\cal B})}
\]
{{\it such that}}\\

{\indent (1) $h^{0} = h_{M}:~ {\cal A}(M) ~{\longrightarrow}~ {\cal B}(M)$}\\

{\indent (2) $h^{q} = {\mbox{\it Identity}}$ {\it if} $h = {\mbox{\it
Identity}}, q ~{\geq}~ 0$}\\

{\indent (3) $g^{q} ~{\bf {\circ}}~ h^{q} = (g ~{\bf {\circ}}~ h)^{q},
{}~{\forall} ~q
{}~{\geq}~ 0$, {\it if}~ $g:~ {\cal B} ~{\longrightarrow}~ {\cal C}$ {\it is
another sheaf
morphism.}}
\\

{\noindent {\it (c) For each short exact sequence of sheaves}}
\[
{1 ~{\longrightarrow}~ {\cal A} ~{\longrightarrow}~ {\cal B}
{}~{\longrightarrow}~ {\cal
C} ~{\longrightarrow}~ 1}
\]
{{\it there is a group homomorphism ${\delta}^{q}:~ H^{q}(M; {\cal C})
{}~{\longrightarrow}~
H^{q ~+~ 1}(M; {\cal A}), ~{\forall} ~q ~{\geq}~ 0$ such that}}\\

{\noindent {\it (1) The induced sequence}}
\[
\begin{array}{cl}
1 &{\longrightarrow}~ H^{0}(M; {\cal A}) ~{\longrightarrow}~ H^{0}(M; {\cal B})
\\
 &{\longrightarrow}~ H^{0}(M; {\cal C})
{}~{\stackrel{\delta^{1}}{\longrightarrow}}~
H^{1}(M; {\cal A}) ~{\longrightarrow}~ ... \\
 &{\longrightarrow}~ H^{q}(M; {\cal A}) ~{\longrightarrow}~ H^{q}(M; {\cal B})
{}~{\longrightarrow}~ H^{q}(M; {\cal C}) \\
 &{\stackrel{\delta^{q}}{\longrightarrow}}~ H^{q ~+~ 1}(M; {\cal A})
{}~{\longrightarrow}~ ~~.~~.~~.
\end{array}
\]
{{\it is {\underline{exact}}.}}\\
\\
{\noindent {\it (2) A commutative diagram}}
\[
\begin{array}{ccccccccc}
1 &{\longrightarrow} &{\cal A} &{\longrightarrow} &{\cal B} &{\longrightarrow}
&{\cal
C} &{\longrightarrow} &1 \\
 & &{\downarrow} & &{\downarrow} & &{\downarrow} &  & \\
1 &{\longrightarrow} &{\cal A}^{1} &{\longrightarrow} &{\cal B}^{1}
&{\longrightarrow}
&{\cal C}^{1} &{\longrightarrow} &1
\end{array}
\]

{\noindent {\it induces a commutative diagram}}
\[
\begin{array}{ccccccc}
1 &{\longrightarrow} &H^{0}(M; {\cal A}) &{\longrightarrow} &H^{0}(M; {\cal B})
&{\longrightarrow} &H^{0}(M; {\cal C}) \\
 & &{\downarrow} & &{\downarrow} & &{\downarrow} \\
1 &{\longrightarrow} &H^{0}(M; {\cal A}^{1}) &{\longrightarrow} &H^{0}(M; {\cal
B}^{1}) &{\longrightarrow} &H^{0}(M; {\cal C}^{1}) \\
 & & & & & & \\
 &{\longrightarrow} &H^{1}(M; {\cal A}) &{\longrightarrow} &H^{1}(M; {\cal B})
&{\longrightarrow} &H^{1}(M; {\cal C}) \\
 & &{\downarrow} & &{\downarrow} & &{\downarrow} \\
 &{\longrightarrow} &H^{1}(M; {\cal A}^{1}) &{\longrightarrow} &H^{1}(M; {\cal
B}^{1}
&{\longrightarrow} &H^{1}(M; {\cal C}^{1}) \\
 & & & & & & \\
 &{\longrightarrow} & & & & & \\
 & & & & & & \\
 &{\longrightarrow} &.~~.~~. & & & &
\end{array}
\]
\\

{\noindent {\it Proof}~ Wells [5], page 57.}\\
\\
{The `connecting homomorphisms', ${\delta}^{q}$, are known as {\it Bockstein
homomorphisms}, and will play a crucial role in our discussion in the next
section.}\\
\vspace*{0.6cm}

{\noindent \bf III. ~Derivation of the obstructions to non-Cliffordian pin
structures}\\

{First, let us adopt the shorthand $P = {\mbox{Pin}}^{a, b, c}(p, q), ~P_{0} =
{\mbox{Pin}}_{0}^{a, b, c}(p, q), ~O_{0}$  $= O_{0}(p, q), ~O = O(p, q), ~C =
C^{a, b, c}$ in order to more efficiently describe the groups of Section I;
associated
to these groups are then constant sheaves ${\cal P}, {\cal P}_{0}, {\cal
O}_{0},
{\cal O}$, and ${\cal C}$. Associated to diagram (3), then, is the following
commutative diagram of sheaf morphisms:}
\[
\begin{array}{cccccccccc}
 & &| & &| & &| & & & \\
 & &{\downarrow} & &{\downarrow} & &{\downarrow} & & & \\
 & &{\Bbb Z_{2}} & &{\Bbb Z_{2}} & &{\Bbb Z_{2}} & & & \\
 & &{\downarrow} & &{\downarrow} & &{\downarrow} & & & \\
1 &{\longrightarrow} &{\cal P}_{0} &{\longrightarrow} &{\cal P}
&{\longrightarrow}
&{\cal C} &{\longrightarrow} &1 & \\
 & &{\downarrow} & &{\downarrow} & &{\downarrow} & & &\hspace*{3.1cm}
{}~~~~~~~~\hfill
(5) \\
1 &{\longrightarrow} &{\cal O}_{0} &{\longrightarrow} &{\cal O}
&{\longrightarrow}
&{\Bbb Z_{2}} ~{\times}~ {\Bbb Z_{2}} &{\longrightarrow} &1 & \\
 & &{\downarrow} & &{\downarrow} & &{\downarrow} & & & \\
 & &| & &| & &| & & &
\end{array}
\]

{\noindent where the horizontal and vertical sequences are all exact. Combining
diagram (5) with the above Theorem, we obtain the following commutative diagram
of
sheaf cohomology groups:}
\[
\begin{array}{cccccccc}
 & &| & &| & &| & \\
 & &{\downarrow} & &{\downarrow} & &{\downarrow} & \\
 & &H^{0}(M; {\Bbb Z_{2}}) & &H^{0}(M; {\Bbb Z_{2}}) & &H^{0}(M; {\Bbb Z_{2}})
& \\
 & & & &{\downarrow} & &{\downarrow} & \\
1 &{\longrightarrow} &H^{0}(M; {\cal P}_{0}) &{\longrightarrow} &H^{0}(M; {\cal
P})
&{\longrightarrow} &H^{0}(M; {\cal C}) & \\
 & &{\downarrow} & &{\downarrow} & &{\downarrow} & \\
1 &{\longrightarrow} &H^{0}(M; {\cal O}_{0}) &{\longrightarrow} &H^{0}(M; {\cal
O})
&{\longrightarrow} &H^{0}(M; {\Bbb Z_{2}} ~{\times}~ {\Bbb Z_{2}}) & \\
 & &{\downarrow} & &{\downarrow} & &{\downarrow} & \\
 & &H^{1}(M; {\Bbb Z_{2}}) & &H^{1}(M; {\Bbb Z_{2}}) & &H^{1}(M; {\Bbb Z_{2}})
& \\
 & &{\downarrow} & &{\downarrow} & &{\downarrow}f &\hspace*{0.55cm}
{}~~~~~~\hfill (6) \\
 &{\longrightarrow} &H^{1}(M; {\cal P}_{0}) &{\longrightarrow} &H^{1}(M; {\cal
P})
&{\stackrel{\longrightarrow}{\tilde p}} &H^{1}(M; {\cal C}) & \\
 & &{\downarrow} & &{\alpha}{\downarrow} & &{\downarrow}{\beta} & \\
 &{\longrightarrow} &H^{1}(M; {\cal O}_{0}) &{\longrightarrow} &H^{1}(M; {\cal
O})
&{\stackrel{\longrightarrow}{p}} &H^{1}(M; {\Bbb Z_{2}} ~{\times}~ {\Bbb
Z_{2}}) & \\
 & &{\downarrow} & &{\delta}_{0}^{2}{\downarrow} & &{\delta}_{2}{\downarrow} &
\\
 & &H^{2}(M; {\Bbb Z_{2}}) & &H^{2}(M; {\Bbb Z_{2}}) & &H^{2}(M; {\Bbb Z_{2}})
&
\end{array}
\]
\\
{\indent We are interested in the bottom part of this diagram (where we have
labelled
the maps between cohomology groups). Recalling that the vertical sequences in
this
diagram are exact, the derivation of the obstructions proceeds as follows.

Let ${\xi} ~{\in}~ H^{1}(M; {\cal O})$, i.e., ${\xi}$ is a principal $O(p,
q)$-bundle
over $M$. We are concerned with the obstruction to the existence of a principal
Pin$(p, q)$ bundle, ${\tilde {\xi}} ~{\in}~ H^{1}(M; {\cal P})$, over ${\xi}$.

Thus, suppose that such a Pin$(p, q)$ bundle, ${\tilde {\xi}}$, exists. Then
${\alpha}({\tilde {\xi}}) ~{\in}~ H^{1}(M;$  ${\cal O})$, and so by exactness}
\[
{{\delta}_{0}^{2}({\alpha}({\tilde {\xi}})) = 0 .}
\]

{That is, if ${\alpha}({\tilde {\xi}}) = {\xi}$, then we must have that
$H^{2}(M;
{\Bbb Z_{2}}) ~{\ni}~ {\delta}_{0}^{2}({\xi}) = 0$. Likewise, if
${\delta}_{0}^{2}({\xi}) =
0$, then such a ${\tilde {\xi}} ~{\in}~ H^{1}(M; {\cal P})$ exists, and so we
see that
the obstruction to the existence of a Pin$(p, q)$ bundle ${\tilde {\xi}}$ is
the
vanishing of the class ${\delta}_{0}^{2}({\xi}) ~{\in}~ H^{2}(M; {\Bbb Z_{2}})$
(here
we are regarding ${\Bbb Z_{2}}$ additively, i.e., ${\Bbb Z_{2}} = {\{}0,
1{\}}$).

The point now is that we can `transfer' the above argument over to the vertical
exact
sequence on the far right in diagram (6). In other words, if ${\tilde {\xi}}
{}~{\in}~
H^{1}(M; {\cal P})$ exists over $M$, then by the commutativity of (6),}
\[
{{\beta}({\tilde p}({\tilde {\xi}})) = p({\alpha}({\tilde {\xi}})) ~{\in}~
H^{1}(M;
{\Bbb Z_{2}} ~{\times}~ {\Bbb Z_{2}}),}
\]
{and so the obstruction is now}
\[
{{\delta}_{2}({\beta}({\tilde p}({\tilde {\xi}}))) =
{\delta}_{2}(p({\alpha}({\tilde
{\xi}}))) ~{\in}~ H^{2}(M; {\Bbb Z_{2}}).}
\]

{Now, by Milnor and Stasheff [8] the general form of this obstruction must be}
\[
{H_{2}(M; {\Bbb Z_{2}}) ~{\ni}~ w_{2}({\tau}_{M}) ~+~ w_{1}({\tau}_{M})
{}~{\smile}~
w_{1}({\tau}_{M}) \hspace*{4.45cm} ~~~~~~~\hfill (7)}
\]
{where $w_{1}({\tau}_{M})$ and $w_{2}({\tau}_{M})$ are the first and second
Stiefel-Whitney classes of ${\tau}_{M}$, respectively.

Decomposing the tangent bundle ${\tau}_{M}$ as}
\[
{{\tau}_{M} ~{\simeq}~ {\tau}^{+} ~{\oplus}~ {\tau}^{-}}
\]
{(where the `plus' and `minus' signs of the subbundles refer to the behaviour
of
sections of these bundles with respect to the Lorentz metric) we obtain}
\[
{w_{1}({\tau}_{M}) = w_{1}({\tau}^{+} ~{\oplus}~ {\tau}^{-}) =
w_{1}({\tau}^{+})
{}~+~ w_{1}({\tau}^{-}) \hspace*{4.00cm} ~~~~~~~~\hfill (8)}
\]
{and}
\[
{w_{2}({\tau}_{M}) = w_{2}({\tau}^{+}) ~+~ w_{2}({\tau}^{-}) ~+~
w_{1}({\tau}^{+})
{}~{\smile}~ w_{1}({\tau}^{-}) \hspace*{2.65cm} ~~~~~~~~~~\hfill (9)}
\]

{Combining (8) and (9), and adopting the conventions $w_{1}({\tau}^{+}) =
w_{1}^{+},
w_{1}({\tau}^{-})$ $= w_{1}^{-}, w_{2}({\tau}^{+}) = w_{2}^{+},
w_{2}({\tau}^{-}) =
w_{2}^{-}$ we see that the obstruction must have the general form}
\[
\begin{array}{cc}
aw_{2}^{+} ~+~ bw_{2}^{-} ~+~ cw_{1}^{-} ~{\smile}~ w_{1}^{-} ~+~ dw_{1}^{+}
{}~{\smile}~
w_{1}^{-} ~+~ ew_{1}^{+} ~{\smile}~ w_{1}^{+} & \\
= {\delta}_{2}(p({\xi})) ~{\in}~ H^{2}(M; {\Bbb Z_{2}}) &\hspace*{0.8cm}
{}~~~~~~~~\hfill (10)
\end{array}
\]
{where $a, b, c, d, e ~{\in}~ {\Bbb Z_{2}}$ are constants yet to be determined.
Clearly,
then, the determination of $a, b, c, d$, and $e$ depends upon the nature of the
double
cover given by the exact sequence}
\[
{1 ~{\longrightarrow}~ {\Bbb Z_{2}} ~{\longrightarrow}~ C^{a, b, c}
{}~{\longrightarrow}~
{\Bbb Z_{2}} ~{\times}~ {\Bbb Z_{2}} ~{\longrightarrow}~ 1 \hspace*{4.8cm}
{}~~~~~~~~\hfill (11)}
\]
{that is to say, the values of $a, b, c, d, e ~{\in}~ {\Bbb Z_{2}}$ depend upon
the
choice of $C^{a, b, c}$. We will treat each of these choices in turn. First,
however,
we need to understand the `Bockstein' homomorphism, ${\delta}_{2}:~ H^{1}(M;
{\Bbb Z_{2}} ~{\times}~ {\Bbb Z_{2}}) ~{\longrightarrow}~ H^{2}(M; {\Bbb
Z_{2}})$:

To begin, recall the interpretation of $w_{1}^{+}$ and $w_{1}^{-}$:

Once we have decomposed the tangent bundle ${\tau}_{M}$ via ${\tau}_{M}
{}~{\simeq}~
{\tau}^{+} ~{\oplus}~ {\tau}^{-}$, we have the notions of
{\it time-orientability} and {\it space-orientability} [9]. Then
$w_{1}^{+}$ and $w_{1}^{-}$ are cohomological data which tell us about the
orientation of $M$. For example, if the signature is $(- + + +)$ then (i)
$w_{1}^{-} =
0 ~{\Longleftrightarrow}~ M$ is time-orientable, and (ii) $w_{1}^{+} = 0
{}~{\Longleftrightarrow}~ M$ is space-orientable.

More formally, what this means is that $w_{1}^{+}$ and $w_{1}^{-}$ define a
${\Bbb
Z_{2}} ~{\times}~ {\Bbb Z_{2}}$ valued ${\breve C}$ech 1-cochain,}
\[
{(w_{1}^{+}, w_{1}^{-}):~ U_{a} ~{\cap}~ U_{b} ~{\longrightarrow}~ {\Bbb Z_{2}}
{}~{\times}~ {\Bbb Z_{2}}}
\]
{(where $U_{a}$ and $U_{b}$ are two non-empty open sets in some arbitrary
simple cover
of $M$). In other words, $(w_{1}^{+}, w_{1}^{-}) ~{\in}~ H^{1}(M; {\Bbb Z_{2}}
{}~{\times}~
{\Bbb Z_{2}})$. We therefore expect the Bockstein homomorphism, ${\delta}_{2}$,
to
relate the elements $(w_{1}^{+}, w_{1}^{-}) ~{\in}~ H^{1}(M; {\Bbb Z_{2}}
{}~{\times}~$
${\Bbb Z_{2}})$ to the elements in $w_{1}^{+} ~{\smile}~ w_{1}^{+}, w_{1}^{-}
{}~{\smile}~ w_{1}^{-}$, etc. in $H^{2}(M; {\Bbb Z_{2}})$. To see how this
occurs, recall
the formal definition of ${\delta}_{2}$ [5].

First, consider the following commutative diagram of exact sequences:}
\[
\begin{array}{cccccccccc}
1 &{\longrightarrow} &C^{2}({\Bbb Z_{2}}) &{\stackrel{f^{1}}{\longrightarrow}}
&C^{2}({\cal C}) &{\stackrel{g^{1}}{\longrightarrow}} &C^{2}({\Bbb Z_{2}}
{}~{\times}~
{\Bbb Z_{2}}) &{\longrightarrow} &1 & \\
 & &{\uparrow}{\alpha} & &{\uparrow}{\beta} & &{\uparrow}{\gamma} & &
& \hspace*{2.1cm} \hfill (12)\\
1 &{\longrightarrow} &C^{1}({\Bbb Z_{2}}) &{\stackrel{f}{\longrightarrow}}
&C^{1}({\cal C}) &{\stackrel{g}{\longrightarrow}} &C^{1}({\Bbb Z_{2}}
{}~{\times}~
{\Bbb Z_{2}}) &{\longrightarrow} &1 &
\end{array}
\]
{where $C^{n}({\cal A})$ is the set of $n$-cochains with coefficients in ${\cal
A}$.

Let $c ~{\in}~ {\mbox{ker}}({\gamma})$. Then $c = g(c^{1})$, for some $c^{1}
{}~{\in}~
C^{1}({\cal C})$, by exactness ($c$ gives us a cohomology class in $C^{1}({\Bbb
Z_{2}}
{}~{\times}~ {\Bbb Z_{2}})$). By commutativity, we get $g^{1}({\beta}(c^{1})) =
{\gamma}(g(c^{1})) = 1$, and so ${\beta}(c^{1}) = f^{1}(a)$, for some $a
{}~{\in}~
C^{2}({\Bbb Z_{2}})$. We then get an induced mapping}
\[
{{\delta}_{2}:~ H^{1}(M; {\Bbb Z_{2}} ~{\times}~ {\Bbb Z_{2}})
{}~{\longrightarrow}~
H^{2}(M; {\Bbb Z_{2}})}
\]
{from the map given above,}
\[
{C^{1}({\Bbb Z_{2}} ~{\times}~ {\Bbb Z_{2}}) ~{\ni}~ c ~{\longrightarrow}~
(f^{1})^{-1} ~{\bf
{\circ}}~ {\beta}(g^{-1}(c)) = a ~{\in}~ C^{2}({\Bbb Z_{2}}) .}
\]
{Since the homomorphism inducing $g$ depends on the choice of $C^{a, b, c}$, we
see
that ${\delta}_{2}$ depends upon our choice of $C^{a, b, c}$.

In fact, the above construction shows us how to calculate the images of
$(w_{1}^{+},
0), (0, w_{1}^{-}) ~{\in}~ H^{1}(M; {\Bbb Z_{2}} ~{\times}~ {\Bbb Z_{2}})$ in
$H^{2}(M; {\Bbb Z_{2}})$ under ${\delta}_{2}$. For example, if we take the
signature to
be $(- + + +)$ then $(w_{1}^{+}, 0)$ and $(0, w_{1}^{-})$ are related to the
transformations $(R_{S}, 0)$ and $(0, R_{T})$ in the obvious way, i.e.,
$(w_{1}^{+}, 0)$ tells us whether or not we can continuously distinguish
between
systems under the operation $(R_{S}, 0)$, and likewise for time reversal. Now,
the
elements $(R_{S}, 0)$ and $(0, R_{T})$ are double covered by elements
${\pm}{\tilde R_{S}}$ and ${\pm}{\tilde R_{T}}$ (respectively) in $C^{a, b,
c}$.
Corresponding to the way the elements $(R_{S}, 0), (0, R_{T}) ~{\in}~ {\Bbb
Z_{2}}
{}~{\times}~ {\Bbb Z_{2}}$ are covered by elements in $C^{a, b, c}$, there is
also a
`lifting' of the elements $(w_{1}^{+}, 0), (0, w_{1}^{-}) ~{\in}~ H^{1}(M;
{\Bbb
Z_{2}} ~{\times}~ {\Bbb Z_{2}})$ to elements ${\pm}{\tilde w_{+}}, {\pm}{\tilde
w_{-}}
{}~{\in}~ H^{1}(M; {\cal C})$ (corresponding to the map $g^{-1}$in (12) above).
Next, we
apply the Steenrod square operation $Sq^{1}$(corresponding to the map
${\beta}$in
(12)), i.e.,}
\[
{Sq^{1}({\tilde w_{1}^{\pm}}) = {\tilde w_{1}^{\pm}} ~{\smile}~ {\tilde
w_{1}^{\pm}}
{}~{\in}~ C^{2}({\cal C}) .}
\]
{Finally, we pull the elements $Sq^{1}({\tilde w_{\pm}})$to elements
$w_{1}^{\pm}
{}~{\smile}~ w_{1}^{\pm} ~{\in}~ H^{2}(M; {\Bbb Z_{2}})$ (corresponding to the
map
$(f^{1})^{-1}$). The point is, when we pulled back $(w_{1}^{+}, 0)$ (say) to
${\tilde
w_{1}^{+}} ~{\in}~ C^{1}({\cal C})$, we did so in a way compatible with the
homomorphism $C^{a, b, c} ~{\stackrel{f^{*}}{\longrightarrow}}~ {\Bbb Z_{2}}
{}~{\times}~ {\Bbb Z_{2}}$, i.e., if the 1-cycle, $c_{1}$, dual to $(w_{1}^{+},
0)$ satisfies $<(w_{1}^{+}, 0), c_{1}> = a$, then the 1-cycle, $c_{1}$, dual to
${\tilde w_{1}^{+}}$ must satisfy $<{\tilde w_{1}^{+}}, c_{1}^{\prime}> =
{\tilde a}$,
where ${\pm}{\tilde a}$ covers $a$ under the homomorphism $f^{*}$. When we then
apply
$Sq^{1}$to ${\tilde w_{1}^{+}}$we obtain ${\tilde w_{1}^{+}} ~{\smile}~ {\tilde
w_{1}^{+}}$, with the property that for some 2-cycle, $c_{2}$, dual to ${\tilde
w_{1}^{+}} ~{\smile}~ {\tilde w_{1}^{+}}$ we have $<{\tilde w_{1}^{+}}
{}~{\smile}~
{\tilde w_{1}^{+}}, c_{2}> = <{\tilde w_{1}^{+}}, {\mbox{`front 1-face of
$c_{2}$'}}>
{\bf {\cdot}} <{\tilde w_{1}^{+}}, {\mbox{`back 1-face of $c_{2}$'}}> = {\tilde
a^{2}}$. In other words, the pull back of ${\tilde w_{1}^{+}} ~{\smile}~
{\tilde
w_{1}^{+}} ~{\in}~ C^{2}$ ~$({\cal C})$ to ${\delta}_{2}(w_{1}^{+}, 0) ~{\in}~
H^{2}(M;
{\Bbb Z_{2}})$ will depend upon whether or not ${\tilde a^{2}} ~{\in}~ C^{a, b,
c}$
pulls back to 0 or 1 in the group ${\Bbb Z_{2}}$ under the homomorphism
$f^{*}$. If
${\tilde a^{2}}$ pulls back to 0, then ${\delta}_{2}(w_{1}^{+}) = 0$.
Otherwise,
$e = 1$.

Furthermore, we see from the above construction that the class
$w_{2}({\tau}_{M}) =
w_{2}^{+} ~+~ w_{2}^{-} ~+~ w_{1}^{+} ~{\smile}~ w_{1}^{-}$ is unaffected by
the
choice of $C^{a, b, c}$, i.e., we always have $a = b = d = 1$.

We now proceed with a case by case analysis.}\\
\\
{\noindent \bf ${\boldmath C^{a, b, c} ~{\simeq}~ {\Bbb Z_{2}} ~{\times}~ {\Bbb
Z_{2}}
{}~{\times}~ {\Bbb Z_{2}}}$}\\

{Recall that taking $C^{a, b, c} ~{\simeq}~ {\Bbb Z_{2}} ~{\times}~ {\Bbb
Z_{2}}
{}~{\times}~ {\Bbb Z_{2}}$ is equivalent (in Dabrowski's notation) to
considering the
groups ${\mbox{Pin}}^{+, +, +}(p, q)$. We are then concerned with seeing how
$(w_{1}^{+}, w_{1}^{-}) ~{\in}~ H^{1}(M; {\Bbb Z_{2}} ~{\times}~ {\Bbb Z_{2}})$
`pulls
back' under the sequence of homomorphisms}
\[
{H^{1}(M; {\Bbb Z_{2}}) ~{\stackrel{f}{\longrightarrow}}~ H^{1}(M; {\Bbb Z_{2}}
{}~{\times}~ {\Bbb Z_{2}} ~{\times}~ {\Bbb Z_{2}})
{}~{\stackrel{g}{\longrightarrow}}~
H^{1}(M; {\Bbb Z_{2}} ~{\times}~ {\Bbb Z_{2}})}
\]
{induced by the exact sequence of homomorphisms ${\Bbb Z_{2}}
{}~{\stackrel{f_{*}}
{\longrightarrow}}~ {\Bbb Z_{2}} ~{\times}~ {\Bbb Z_{2}} ~{\times}~ {\Bbb
Z_{2}}
{}~{\stackrel{g_{*}}{\longrightarrow}}~ {\Bbb Z_{2}} ~{\times}~ {\Bbb Z_{2}}$.
Recall,
however, that the homomorphisms $f_{*}$ and $g_{*}$ can be given explicitly as
shown here in this example (signature $(- + + +)$):}
\[
\begin{array}{ccccc}
{\underline{{\Bbb Z_{2}}}} &{\stackrel{f_{*}}{\longrightarrow}}
&{\underline{{\Bbb Z_{2}} ~{\times}~ {\Bbb Z_{2}} ~{\times}~ {\Bbb Z_{2}}}}
&{\stackrel{\beta_{*}}{\longrightarrow}} &{\underline{{\Bbb Z_{2}} ~{\times}~
{\Bbb Z_{2}}}} \\
 & & & & \\
\begin{array}{c}
0 \\
1
\end{array} & \begin{array}{c}
{\longrightarrow} \\
{\longrightarrow}
\end{array} & \left. \begin{array}{c}
(0, 0, 0) \\
(0, 0, 1)
\end{array}
\right\} &{\longrightarrow} &(0, 0) \\
 & & & & \\
b = + & &\left. \begin{array}{c}
(0, 1, 0) \\
(0, 1, 1)
\end{array}
\right\} &{\longrightarrow} &(0, R_{T}) \\
 & & & & \\
a = + & &\left. \begin{array}{c}
(1, 0, 0) \\
(1, 0, 1)
\end{array}
\right\} &{\longrightarrow} &(R_{S}, 0) \\
 & & & & \\
c = + & &\left. \begin{array}{c}
(1, 1, 0) \\
(1, 1, 1)
\end{array}
\right\} &{\longrightarrow} &(R_{S}, R_{T})
\end{array}
\]
\\
{\indent Now, since the squares of all the elements covering $(R_{S}, 0), (0,
R_{T})$,
and $(R_{S}, R_{T})$ are always $(0, 0, 0) = {\mbox{`the identity in ${\Bbb
Z_{2}}
{}~{\times}~ {\Bbb Z_{2}} ~{\times}~ {\Bbb Z_{2}}$'}}$, we see that we can
always pull
back the elements $(w_{1}^{+}, 0), (0, w_{1}^{-}) ~{\in}~ H^{1}(M; {\Bbb Z_{2}}
{}~{\times}~ {\Bbb Z_{2}})$ to elements ${\tilde w_{1}^{+}}, {\tilde w_{1}^{-}}
{}~{\in}~
H^{1}(M; {\cal C})$ with the property that ${\tilde w_{1}^{+}} ~{\smile}~
{\tilde
w_{1}^{+}}, {\tilde w_{1}^{-}} ~{\smile}~ {\tilde w_{1}^{-}} ~{\in}~ H^{2}($
{}~$M; {\cal
C})$ are both zero cocycles. Thus, pulling these cocycles back under $f$
(induced by
$f^{*}$ given above) we get}
\[
{{\delta}_{2}(w_{1}^{+}, 0) = {\delta}_{2}(0, w_{1}^{-}) = 0 ~{\in}~ H^{2}(M;
{\Bbb
Z_{2}}) .}
\]
{In other words, $c = e = 0$, and so the information contained in $(w_{1}^{+},
0)$ and
$(0, w_{1}^{-})$ is not relevant to the obstruction class in this situation.
Thus, we
have shown}\\
\\
{\noindent {\bf Theorem 1.}~{\it Let $M$ be a spacetime with tangent bundle
${\tau}_{M}$ an $O(p, q)$ bundle. Then $M$ admits ${\mbox{Pin}}^{+, +, +}(p,
q)$
structure if and only if}}
\[
{w_{2}^{+} ~+~ w_{2}^{-} ~+~ w_{1}^{-} ~{\smile}~ w_{1}^{+} = 0}
\]
{{\it where $w_{2}^{\pm}$, $w_{1}^{\pm}$ are defined as above.}}\\
\\
{\noindent \bf ${\boldmath C^{a, b, c} ~{\simeq}~ D_{4}}$}\\

{Recall that taking $C^{a, b, c} ~{\simeq}~ D_{4}$ yields the
{\it Cliffordian} pin groups ${\mbox{Pin}}^{+, -, +}(p,$  $q)$ and
${\mbox{Pin}}^{-, +, +}(p, q)$. Although the obstructions to these structures
have
been worked out [3], we present our approach here for completeness.

Thus, recall that $D_{4}$ can be regarded as a semidirect product, $D_{4}
{}~{\simeq}~
{\Bbb Z_{4}} ~{\odot}~ {\Bbb Z_{2}}$, where ${\Bbb Z_{4}} ~{\subset}~ D_{4}$ is
a
normal subgroup, i.e., elements $(a_{1}, b_{1}), (a_{2}, b_{2}) ~{\in}~ D_{4}
{}~{\simeq}~ {\Bbb Z_{4}}$ $~{\odot}~ {\Bbb Z_{2}}$ multiply according to
$(a_{1}, b_{1})
{}~{\bf {\cdot}}~ (a_{2}, b_{2}) = (a_{1} b_{1} a_{2} b_{1}^{-1}, b_{1}
b_{2})$. If we
regard $a$ as the generator of the `${\Bbb Z_{4}}$ part' ($a^{4} = 0$) and $b$
as the
generator of the `$Z_{2}$ part' ($b^{2} = 0$), then what this means is that
there are
two different cases, corresponding to either the groups ${\mbox{Pin}}^{+, -,
+}(1, 3)$
and ${\mbox{Pin}}^{-, +, +}(3, 1)$ or the groups ${\mbox{Pin}}^{+, -, +}(3, 1)$
and
${\mbox{Pin}}^{-, +, +}(1, 3)$. For the group ${\mbox{Pin}}^{+, -, +}(1, 3)$ we
get the
sequence of homomorphisms}
\[
\begin{array}{ccccc}
{\underline{{\Bbb Z_{2}}}} &{\stackrel{f_{*}}{\longrightarrow}}
&{\underline{D_{4}}}
&{\stackrel{\beta_{*}}{\longrightarrow}} &{\underline{{\Bbb Z_{2}} ~{\times}~
{\Bbb Z_{2}}}} \\
 & & & & \\
\begin{array}{c}
0 \\
1
\end{array} & \begin{array}{c}
{\longrightarrow} \\
{\longrightarrow}
\end{array} &\left. \begin{array}{c}
(0, 0) \\
(a^{2}, 0)
\end{array}
\right\} &{\longrightarrow} &(0, 0) \\
 & & & & \\
a = + & &\left. \begin{array}{c}
(a, b) \\
(a^{3}, b)
\end{array}
\right\} &{\longrightarrow} &(0, R_{S}) \\
 & & & & \\
b = - & &\left. \begin{array}{c}
(a, 0) \\
(a^{3}, 0)
\end{array}
\right\} &{\longrightarrow} &(R_{T}, 0) \\
 & & & & \\
c = + & &\left. \begin{array}{c}
(0, b) \\
(a^{2}, b)
\end{array}
\right\} &{\longrightarrow} &(R_{T}, R_{S})
\end{array}
\]
\\
{\indent Now, note the elements covering $(R_{T}, 0), (a, 0)$ and $(a^{3}, 0)$,
both
satisfy $(a, 0) ~{\bf {\cdot}}~ (a, 0) = a^{2} = (a^{3}, 0) ~{\bf {\cdot}}~
(a^{3},
0)$, i.e., their squares are not equal to the identity element $(0, 0) ~{\in}~
D_{4}$.
It follows that $(w_{1}^{+}, 0)$ pulls back to $w_{1}^{+} ~{\smile}~
w_{1}^{+}$, i.e.,
${\delta}_{2}(w_{1}^{+}, 0) = w_{1}^{+} ~{\smile}~ w_{1}^{+}$ and so $e = 1, c
= 0$.
Thus, we have shown}\\
\\
{\noindent {\bf Theorem 2.}~{\it Let $M$ be a spacetime with tangent bundle
${\tau}_{M}$ either an $O(3, 1)$ bundle or an $O(1, 3)$ bundle; then $M$ admits
either
${\mbox{Pin}}^{-, +, +}(3, 1)$ or ${\mbox{Pin}}^{+, -, +}(1, 3)$ structure
(respectively) if and only if}}
\[
{w_{2}^{+} ~+~ w_{2}^{-} ~+~ w_{1}^{+} ~{\smile}~ w_{1}^{-} ~+~ w_{1}^{+}
{}~{\smile}~
w_{1}^{+} = 0 .}
\]

{When we consider the sequence of homomorphisms corresponding to the groups
${\mbox{Pin}}^{+, -, +}(3, 1)$ and ${\mbox{Pin}}^{-, +, +}(1, 3)$, we see that
now it
is $(0, w_{1}^{-})$ that pulls back, and so $c = 1, e = 0$.}\\
\\
{\noindent {\bf Theorem 3.}~{\it Let $M$ be a spacetime with tangent bundle
${\tau}_{M}$ either an $O(3, 1)$ bundle or an $O(1, 3)$ bundle; then $M$ admits
either
${\mbox{Pin}}^{+, -, +}(3, 1)$ or ${\mbox{Pin}}^{-, +, +}(1, 3)$ structure
(respectively) if and only if}}
\[
{w_{2}^{+} ~+~ w_{2}^{-} ~+~ w_{1}^{+} ~{\smile}~ w_{1}^{-} ~+~ w_{1}^{-}
{}~{\smile}~
w_{1}^{-} = 0 .}
\]
\\
{\noindent \bf ${\boldmath C^{a, b, c} ~{\simeq}~ {\Bbb Z_{2}} ~{\times}~
{\Bbb Z_{4}}}$}\\

{Recall that taking $C^{a, b, c} ~{\simeq}~ {\Bbb Z_{2}} ~{\times}~ {\Bbb
Z_{4}}$
corresponds to considering the groups ${\mbox{Pin}}^{a, b, c}(p, q)$, with two
minuses
and one plus occurring in the triple $a, b, c$. Now, we can as usual regard
${\Bbb Z_{2}} ~{\times}~ {\Bbb Z_{4}}$ as the group given abstractly as}
\[
{{\Bbb Z_{4}} ~{\times}~ {\Bbb Z_{2}} ~{\simeq}~ {\{}(a, b)|a^{4} = b^{2} =
1{\}}}
\]
{This means that the homomorphisms associated with the exact sequence (11) are
given,
for the group ${\mbox{Pin}}^{-, -, +}(3, 1)$}
\[
\begin{array}{ccccc}
{\underline{{\Bbb Z_{2}}}} &{\stackrel{f_{*}}{\longrightarrow}}
&{\underline{{\Bbb Z_{4}} ~{\times}~ {\Bbb Z_{2}}}} &{\stackrel{\beta_{*}}
{\longrightarrow}} &{\underline{{\Bbb Z_{2}} ~{\times}~ {\Bbb Z_{2}}}}
\\
 & & & & \\
\begin{array}{c}
0 \\
1
\end{array} & \begin{array}{c}
{\longrightarrow} \\
{\longrightarrow}
\end{array} &\left. \begin{array}{c}
(0, 0) \\
(a^{2}, 0)
\end{array}
\right\} &{\longrightarrow} &(0, 0) \\
 & & & & \\
a = - & &\left. \begin{array}{c}
(a, b) \\
(a^{3}, b)
\end{array}
\right\} &{\longrightarrow} &(R_{S}, 0) \\
 & & & & \\
b = - & &\left. \begin{array}{c}
(a, 0) \\
(a^{3}, 0)
\end{array}
\right\} &{\longrightarrow} &(0, R_{T}) \\
 & & & & \\
c = + & &\left. \begin{array}{c}
(0, b) \\
(a^{2}, b)
\end{array}
\right\} &{\longrightarrow} &(R_{S}, R_{T})
\end{array}
\]
\\
{\indent It follows that both $(w_{1}^{+}, 0)$ and $(0, w_{1}^{-})$ pull back,
and
so $c = e = 1$. Furthermore, this same result clearly holds for the group
${\mbox{Pin}}^{-, -, +}(1, 3)$. Thus, we have shown}\\
\\
{\noindent {\bf Theorem 4.}~{\it Let $M$ be a spacetime with tangent bundle
${\tau}_{M}$ an $O(p, q)$ bundle; then $M$ admits ${\mbox{Pin}}^{-, -, +}(p,
q)$ if
and only if}}
\[
{w_{2}^{+} ~+~ w_{2}^{-} ~+~ w_{1}^{+} ~{\smile}~ w_{1}^{-} ~+~ w_{1}^{+}
{}~{\smile}~
w_{1}^{+} ~+~ w_{1}^{-} ~{\smile}~ w_{1}^{-} = 0}
\]

{For the groups ${\mbox{Pin}}^{+, -, -}(3, 1)$ and ${\mbox{Pin}}^{-, +, -}(1,
3)$, we
see that only $(0, w_{1}^{-})$ pulls back, hence}\\
\\
{\noindent {\bf Theorem 5.}~{\it Let $M$ be a spacetime with tangent bundle
${\tau}_{M}$ either an $O(3, 1)$ bundle or an $O(1, 3)$ bundle; then $M$ admits
either
${\mbox{Pin}}^{+, -, -}(3, 1)$ or ${\mbox{Pin}}^{-, +, -}(1, 3)$ structure
(respectively) if and only if}}
\[
{w_{2}^{+} ~+~ w_{2}^{-} ~+~ w_{1}^{+} ~{\smile}~ w_{1}^{-} ~+~ w_{1}^{-}
{}~{\smile}~
w_{1}^{-} = 0 .}
\]

{Finally, for the remaining cases we obtain}\\
\\
{\noindent {\bf Theorem 6.}~{\it Let $M$ be a spacetime with tangent bundle
${\tau}_{M}$ either an $O(3, 1)$ bundle or an $O(1, 3)$ bundle; then $M$ admits
either
${\mbox{Pin}}^{-, +, -}(3, 1)$ or ${\mbox{Pin}}^{+, -, -}(1, 3)$ structure
(respectively) if and only if}}
\[
{w_{2}^{+} ~+~ w_{2}^{-} ~+~ w_{1}^{+} ~{\smile}~ w_{1}^{-} ~+~ w_{1}^{+}
{}~{\smile}~
w_{1}^{+} = 0 .}
\]
\\
{\noindent \bf ${\boldmath C^{a, b, c} ~{\simeq}~ Q_{4}}$}\\

{Recall that taking $C^{a, b, c} ~{\simeq}~ Q_{4}$ is equivalent to considering
the
groups ${\mbox{Pin}}^{-, -, -}(p, q)$. Clearly then, both $(w_{1}^{+}, 0)$ and
$(0,
w_{1}^{-})$ always pull back. Thus,}\\
\\
{\noindent {\bf Theorem 7.}~{\it Let $M$ be a spacetime with tangent bundle
${\tau}_{M}$ an $O(p, q)$ bundle; then $M$ admits ${\mbox{Pin}}^{-, -, -}(p,
q)$
structure if and only if}}
\[
{w_{1}^{-} ~{\smile}~ w_{1}^{-} ~+~ w_{1}^{+} ~{\smile}~ w_{1}^{+} ~+~
w_{2}^{+} ~+~
w_{2}^{-} ~+~ w_{1}^{+} ~{\smile}~ w_{1}^{-} = 0 .}
\]
\vspace*{0.6cm}

{\noindent \bf IV. ~Applications of the obstructions to pin-Lorentz
cobordism}\\

{In this section, we use the obstructions developed above in Section III to
derive the
obstructions to pin-Lorentz cobordism. First, however, we review some
elementary
concepts from differential topology.

Now, recall that the existence of an everywhere non-singular Lorentz metric on
$M$ is
equivalent to the existence of a global non-vanishing (smooth) {\it line
field}, ${\{}v, -v{\}}$, on $M$ (when $M$ is time-orientable, it suffices that
$M$
possess a global non-vanishing vector field $v$). The vectors ${\pm}v$ then
have the
usual interpretation as timelike vectors (see [9]).

Recall also the notion of {\it kink number}: Let ${\Sigma} ~{\subset}~ M$ be a
three-dimensional, connected submanifold. Since ${\dim}({\Sigma}) = 3$, we can
always
find a global framing ${\{}u_{i}:~ i = 1, 2, 3{\}}$ of ${\Sigma}$. Furthermore,
even
if $M$ is not orientable we can always find a unit line field ${\{}n, -n{\}}$
which is
normal to ${\Sigma}$ (note that $n$ has unit length with respect to the
underlying
Riemannian metric on $M$, $g_{ab}^{R}$, i.e., $g_{ab}^{R}u^{a}u^{b} = 1$). We
can
extend this tetrad framing $(n, u_{i})$ of ${\Sigma}$ to a collar
neighbourhood}
\[
{N ~{\cong}~ {\Sigma} ~{\times}~ [0, 1]}
\]
{(we extend to $N$ to deal with the case ${\Sigma} ~{\cong}~ {\partial}M$). Let
$v$ be
the timelike vector (line field) determined by $g_{ab}$. Then $v$ can be
written as}
\[
{v = v^{0}n ~+~ v^{i}u_{i}}
\]
{such that ${\displaystyle{\sum_{i}}(v^{i})^{2} = 1}$. Clearly, then $v$
determines a
map}
\[
{K:~ {\Sigma} ~{\longrightarrow}~ \left\{ \begin{array}{cl}
S^{3}, &~{\mbox{if $M$ is time-orientable}} \\
 & \\
{\Bbb R}{\Bbb P^{3}}, &~{\mbox{if $M$ is {\it not} time-orientable}}
\end{array}
\right.}
\]
\\
{\noindent by assigning to each point $p ~{\in}~ {\Sigma}$ the direction in
$T_{p}M$
(a point on the $S^{3}$ or ${\Bbb R}{\Bbb P^{3}}$ determined by the tetrad $(n,
u_{i})$) that
$v_{p}$ points to. We then define the kink number of $g_{ab}$ with respect to
${\Sigma}$ by the formula}
\[
{{\mbox{kink}}({\Sigma}; g_{ab}) = {\deg}(K),}
\]
{where ${\deg}(K)$ is `the degree of the mapping $K$'. If $v$ is a timelike
vector
determined by $g_{ab}$, we shall often write}
\[
{{\mbox{kink}}({\Sigma}; g_{ab}) = {\mbox{kink}}({\Sigma}; v) .}
\]

{For our immediate purposes we shall be concerned with kinking wth respect to
${\partial}M$, the boundary of our spacetime. In particular, we shall be
concerned
with the case $M$ compact, with ${\partial}M ~{\cong}~ {\Sigma}_{0} ~{\cup}~
{\Sigma}_{1} ~{\cup}~ ... ~{\cup}~ {\Sigma}_{n}$, where the ${\Sigma}_{i}$'s
are
closed, connected three-manifolds and `${\cup}$' is the operation of disjoint
union.
We wish to define the quantity ${\mbox{kink}}({\partial}M; g_{ab}) =
{\mbox{kink}}({\Sigma}_{0} ~{\cup}~ {\Sigma}_{1} ~{\cup}~ ... ~{\cup}~
{\Sigma}_{n};
g_{ab})$. On differential topological grounds (see [10]) we see that it makes
sense to
write}
\[
{{\mbox{kink}}({\partial}M; g_{ab}) =
{\displaystyle{\sum_{i}}~{\mbox{kink}}({\Sigma}_{i}; g_{ab})}.}
\]

{Now suppose $v$ is a smooth vector field on $M$ which vanishes on some
discrete set
of points $p_{1}, p_{2}, ... p_{n} ~{\in}~ M$. Associated to each of these
vanishing
points $p_{i}$ is the {\it index of $v$ at $p_{i}$}, which is precisely the
degree of mapping given by ${\frac{v(x)}{{\parallel}v(x){\parallel}}}$, which
takes a
little sphere $s(p_{i})$ about $p_{i}$ into the unit sphere. We write
`${\sum}i_{v}$'
to mean `the sum of the indices of v'. We then have the following formula
[10]:}
\[
{{\sum}i_{v} = e(M) ~+~ {\mbox{kink}}({\partial}M; v)}
\]
{where $e(M)$ is the Euler number of $M$ and ${\mbox{kink}}({\partial}M; v)$ is
as
above. In particular, if $M$ is a spacetime then the timelike line field
${\{}v,
-v{\}}$ is non-vanishing and so ${\sum}i_{v} = 0$, hence,}
\[
{e(M) = -{\mbox{kink}}({\partial}M; g_{ab}) \hspace*{8.3cm} \hfill (13)}
\]

{Now, a direct application of Wu's formula ([1] or [8]) shows the following
identity:
For any $x_{2} ~{\in}~ H^{2}(M; {\Bbb Z_{2}})$,}
\[
{w_{2}(M) ~{\smile}~ x_{2} - (w_{1}(M) ~{\smile}~ w_{1}(M)) ~{\smile}~ x_{2} =
x_{2}
{}~{\smile}~ x_{2} \hspace*{3.4cm} \hfill (14)}
\]

{Writing the intersection pairing as $h:~ H_{2}(M; {\Bbb Z_{2}}) ~{\times}~
H_{2}(M;
{\Bbb Z_{2}}) ~{\longrightarrow}~ {\Bbb Z_{2}}$ (defined explicitly via $h(x,
y) = x
{}~{\bf {\cdot}}~ y = (x_{2} ~{\smile}~ y_{2}) ~{\frown}~ w$, where $x_{2},
y_{2} ~{\in}~
H^{2}(M;$  ${\Bbb Z_{2}})$
satisfy $x_{2} ~{\frown}~ w = x$ and $y_{2} ~{\frown}~ w = y$ where $w ~{\in}~
H_{4}(M; {\Bbb Z_{2}})$ is the fundamental homology class) we recall the
important}\\
\\
{\noindent {\bf Lemma}~ (Milnor and Kervaire, [11], page 517).~ {\it Let $M$ be
a smooth manifold of dimension 4. Let $u({\partial}M)$ (the mod 2 Kervaire
semicharacteristic) be given by}}
\[
{u({\partial}M) = {\dim}_{{\Bbb Z_{2}}}(H_{0}({\partial}M; {\Bbb Z_{2}})
{}~{\oplus}~
H_{1}({\partial}M; {\Bbb Z_{2}})) ~{\bmod 2}}
\]
{{\it Then the rank of the intersection pairing, $h$, satisfies}}
\[
{{\mbox{rank}}(h) = (u({\partial}M) ~+~ e(M)) ~{\bmod 2} .}
\]

{{\it Note}: Actually, our version of the above Lemma differs slightly from
that in [11] in that we allow $M$ to be non-orientable. However, the Lemma is
still
true since Poincar{\'e}-Lefshetz duality still holds in ${\Bbb Z_{2}}$
coefficients for
non-orientable $M$.

{}From the definition of $h$ and equation (14) it follows immediately that
${\mbox{rank}}(h) = 0$ if and only if $w_{2} ~+~ w_{1} ~{\smile}~ w_{1} = 0$.
If $M$
is a spacetime, then the Lemma together with equation (13) then give us}\\
\\
{\noindent {\bf Lemma 1.}~ {\it Let $M$ be a spacetime with tangent bundle
${\tau}_{M}$. Then}}
\[
{w_{2}({\tau}_{M}) ~+~ w_{1}({\tau}_{M}) ~{\smile}~ w_{1}({\tau}_{M}) = 0
{}~{\Longleftrightarrow}}
\]
\[
{(u({\partial}M) ~+~ {\mbox{kink}}({\partial}M; g_{ab})) ~{\bmod 2} = 0 .}
\]

{Combining Lemma 1 with equations (8) and (9) and the above set of Theorems, we
obtain
the following:}\\
\\
{\noindent {\bf Definition}~~ Let ${\Sigma}_{1}, {\Sigma}_{2}, ...
{\Sigma}_{n}$ be a collection of closed three-manifolds. Then we say that there
exists
a ${\mbox{Pin}}^{a, b, c}(p, q)$ cobordism for ${\{}{\Sigma}_{i}:~ i = 1, ...
n{\}}$
if and only if there exists a spacetime $M$ admitting ${\mbox{Pin}}^{a, b,
c}(p, q)$
structure and satisfying}
\[
{{\partial}M ~{\cong}~ {\Sigma}_{1} ~{\cup}~ {\Sigma}_{2} ~{\cup}~ ... ~{\cup}~
{\Sigma}_{n} .}
\]

{In the below Corollaries, ${\{}{\Sigma}_{i}:~ i = 1, ... n{\}}$ always denotes
some
collection of closed three-manifolds.}\\
\\
{\noindent {\bf Corollary 1.}~{\it There exists a ${\mbox{Pin}}^{+, +, +}(p,
q)$ cobordism, $M$, for ${\{}{\Sigma}_{i}:~ i = 1, ... n{\}}$ if and only if
the
following holds:}}
\[
{(u({\partial}M) ~+~ {\mbox{kink}}({\partial}M; g_{ab})) ~{\bmod 2} = 0
{}~{\Longleftrightarrow}}
\]
\[
{w_{1}^{+} ~{\smile}~ w_{1}^{+} ~+~ w_{1}^{-} ~{\smile}~ w_{1}^{-} = 0}
\]
\\
{\noindent {\bf Corollary 2.}~{\it There exists either a ${\mbox{Pin}}^{-, +,
+}(3, 1)$ or a ${\mbox{Pin}}^{+, -, +}(1, 3)$ cobordism $M$ for
${\{}{\Sigma}_{i}:~ i
= 1, ... n{\}}$ if and only if the following holds:}}
\[
{(u({\partial}M) ~+~ {\mbox{kink}}({\partial}M; g_{ab})) ~{\bmod 2} = 0
{}~{\Longleftrightarrow}~ w_{1}^{-} ~{\smile}~ w_{1}^{-} = 0}
\]
\\
{\noindent {\bf Corollary 3.}~{\it There exists either a ${\mbox{Pin}}^{+, -,
+}(3, 1)$ or a ${\mbox{Pin}}^{-, +, +}(1, 3)$ cobordism $M$ for
${\{}{\Sigma}_{i}:~ i
= 1, ... n{\}}$ if and only if the following holds:}}
\[
{(u({\partial}M) ~+~ {\mbox{kink}}({\partial}M; g_{ab})) ~{\bmod 2} = 0
{}~{\Longleftrightarrow}~ w_{1}^{+} ~{\smile}~ w_{1}^{+} = 0}
\]
\\
{\noindent {\bf Corollary 4.}~{\it There exists a ${\mbox{Pin}}^{-, -, +}(p,
q)$ cobordism $M$ for ${\{}{\Sigma}_{i}:~ i = 1, ... n{\}}$ if and only if}}
\[
{(u({\partial}M) ~+~ {\mbox{kink}}({\partial}M; g_{ab})) ~{\bmod 2} = 0}
\]
\\
{\noindent {\bf Corollary 5.}~{\it There exists either a ${\mbox{Pin}}^{+, -,
-}(3, 1)$ or a ${\mbox{Pin}}^{-, +, -}(1, 3)$ cobordism $M$ for
${\{}{\Sigma}_{i}:~ i
= 1, ... n{\}}$ if and only if the following holds:}}
\[
{(u({\partial}M) ~+~ {\mbox{kink}}({\partial}M; g_{ab})) ~{\bmod 2} = 0
{}~{\Longleftrightarrow}~ w_{1}^{+} ~{\smile}~ w_{1}^{+} = 0}
\]
\\
{\noindent {\bf Corollary 6.}~{\it There exists either a ${\mbox{Pin}}^{-, +,
-}(3, 1)$ or a ${\mbox{Pin}}^{+, -, -}(1, 3)$ cobordism $M$ for
${\{}{\Sigma}_{i}:~ i
= 1, ... n{\}}$ if and only if the following holds:}}
\[
{(u({\partial}M) ~+~ {\mbox{kink}}({\partial}M; g_{ab})) ~{\bmod 2} = 0
{}~{\Longleftrightarrow}~ w_{1}^{-} ~{\smile}~ w_{1}^{-} = 0}
\]
\\
{\noindent {\bf Corollary 7.}~{\it There exists a ${\mbox{Pin}}^{-, -, -}(p,
q)$ cobordism $M$ for ${\{}{\Sigma}_{i}:~ i = 1, ... n{\}}$ if and only if}}
\[
{(u({\partial}M) ~+~ {\mbox{kink}}({\partial}M; g_{ab})) ~{\bmod 2} = 0}
\]

{Thus, we see that the topological obstructions to ${\mbox{Pin}}^{a, b, c}(p,
q)$
cobordism depend only upon boundary data (i.e., kink number), the values of $a,
b, c
{}~{\in}~ {\{}{\pm}{\}}$, the choice of signature, and the behaviour of the
1-cocycles
$w_{1}^{\pm}$ under the cup product operation.}\\
\vspace*{0.6cm}

{\noindent \bf V. ~Interpreting the breakdown of pin structure}\\

{We now interpret the breakdown of pin structure on $M$ in two different ways:
First, by
examining the behaviour of pinor fields as we parallelly propagate them around
closed loops in
$M$ and secondly, by examining the behaviour of the determinant of the
world line Dirac operator (the fermion effective action which arises in the
quantization of a point particle possessing world line supersymmetry) in these
situations.

Now, first recall that since we are generically dealing with non-orientable
spacetimes $M$ in
this paper, we automatically have ${\pi}_{1}(M) ~{\not=}~ 0$, i.e., $M$ cannot
be simply
connected. This means that there exist loops (closed curves), ${\gamma}$, in
$M$ with the
property that when we parallelly propagate some tetrad $e^{\alpha}$ around
${\gamma}$ we will
reverse the orientation of $e^{\alpha}$.

Explicitly, suppose that we are given an `initial' tetrad $e^{\alpha}_{(i)}$ at
some point $p
{}~{\in}~ {\gamma}$, and that after we parallel propagate around ${\gamma}$ we
are left with a
`final tetrad' $e^{\beta}_{(f)}$. The two tetrads will then be related by the
equation
$e^{\beta}_{(f)} = e^{\alpha}_{(i)} L^{\beta}_{\alpha}$, where
$L^{\beta}_{\alpha} ~{\in}~ O(p,
q)$ is some general Lorentz transformation (note that $L^{\beta}_{\alpha}$
cannot lie in the
identity connected component, $O_{0}(p, q)$, since the final tetrad will
generically have a
different orientation than the initial one). For example, if $e^{\beta}_{(f)}$
has a different
{\it spacelike} orientation than $e^{\alpha}_{(i)}$, then $L^{\beta}_{\alpha}$
must lie
in $P(O_{0}(p, q))$ (the component of $O(p, q)$ containing parity reversal),
and so on.

Now, we wish to view ${\gamma}$ as the initial (and final) curve in a
continuous family of
curves, ${\{}{\gamma}(v)|v ~{\in}~ [0, 1]{\}}$, which begins and ends at
${\gamma}$, i.e.,
${\gamma}(0) = {\gamma}(1) = {\gamma}$. This family of curves sweeps out a
smooth
2-cycle $T$. Thus, for each $v ~{\in}~ [0, 1]$ we have a curve ${\gamma}(v)$
and for
each ${\gamma}(v)$ we parallel propagate some tetrad $e^{\alpha}_{(i)}(v)$
around ${\gamma}(v)$ to obtain a new tetrad
$e^{\beta}_{(f)}(v)$, related to the old one by $e^{\beta}_{(f)}(v) =
L^{\beta}_{\alpha}(v)
e^{\alpha}_{(i)}(v)$ where $L^{\beta}_{\alpha}(v) ~{\in}~ O(p, q)$ for each
value of $v$.

Clearly then, since ${\gamma}(0) = {\gamma}(1) = {\gamma}$ we must have
$L^{\beta}_{\alpha}(0) = L^{\beta}_{\alpha}(1) = I^{\beta}_{\alpha}$.

Now, consider the elements of some `pin bundle' (covering the bundle of frames)
which `represent'
the tetrads $e^{\alpha}_{(i)}(v)$ (which, we note for completeness, constitute
a smooth field of
tetrads on $T$ as we vary $v$), and write these elements as
${\psi}^{\alpha}_{i}(v)$. Then we can consider the problem of parallel
propagating these initial
`pinor fields' around each ${\gamma}(v)$ to obtain final pinor fields
$({\psi}_{f})$ which are
related to the initial ones (on each curve ${\gamma}(v)$) by some
transformation
${\psi}^{\beta}_{f}(v) = {\tilde L}^{\beta}_{\alpha}(v)
{\psi}^{\alpha}_{i}(v)$, where
${}^{\pm}{\tilde L}^{\beta}_{\alpha}(v) ~{\in}~ {\mbox{Pin}}^{a, b, c}(p, q)$
are the elements of
the pin group ${\mbox{Pin}}^{a, b, c}(p, q)$ covering the corresponding Lorentz
transformations
$L^{\beta}_{\alpha}(v) ~{\in}~ O(p, q)$. The point is, again since we have
${\gamma}(0) =
{\gamma}(1) = {\gamma}$ we expect to have ${\tilde L}^{\beta}_{\alpha}(0) =
{\tilde
I}^{\beta}_{\alpha}$ and ${\tilde L}^{\beta}_{\alpha}(1) = {\tilde
I}^{\beta}_{\alpha}$; however,
if there is a breakdown of pin structure we will have ${\tilde
L}^{\beta}_{\alpha}(0) = +{\tilde
I}^{\beta}_{\alpha}$ but ${\tilde L}^{\beta}_{\alpha}(1) = -{\tilde
I}^{\beta}_{\alpha}$. Now, we
saw above (in Section III) that such an anomaly occurs depending upon the value
of a certain
obstruction class, which in turn depends upon the choice of signature and the
values of $a, b, c,
{}~{\in}~ {\{}+, -{\}}$ (the symmetries of the pinor fields). Because of this,
it is useful to
consider an explicit example in order to have a clear picture of what is going
on.

Thus, let $M$ be a spacetime, with signature $(- + + +)$, which is neither
space nor
time-orientable ($w_{1}^{+} = w_{1}^{-} ~{\not=}~ 0$), and consider the problem
of putting a
{\it Cliffordian} pin structure on $M$, i.e., let the pin group be
${\mbox{Pin}}^{+, -, +}(3, 1)$. Then from Section III (Theorem 3) above we know
that
the obstruction to putting this sort of pin structure on $M$ is that the
following
hold:}
\[
{w_{2}^{+} ~+~ w_{2}^{-} ~+~ w_{1}^{+} ~{\smile}~ w_{1}^{-} ~+~ w_{1}^{-}
{}~{\smile}~
w_{1}^{-} = 0}
\]
{A natural question is, why does $w_{1}^{-} ~{\smile}~ w_{1}^{-}$ contribute to
the
possibility of an anomaly but not $w_{1}^{+} ~{\smile}~ w_{1}^{+}$ ? To see how
to
answer this question assume that there exist 2-cycles, $T$ and $T^{\prime}$,
such that
$w_{1}^{-} ~{\smile}~ w_{1}^{-} ~[T] ~{\not=}~ 0$ and $w_{1}^{+} ~{\smile}~
w_{1}^{+}
{}~[T^{\prime}] ~{\not=}~ 0$ (here we are regarding ${\Bbb Z_{2}}$ additively).
It
follows that there are closed curves, ${\gamma}$ and ${\gamma}^{\prime}$,
embedded in
$T$ and $T^{\prime}$ respectively. with the property that when we parallel
propagate a
tetrad  $e_{(i)}$ around ${\gamma}$ the final tetrad has opposite
time-orientation
(i.e., assume for simplicity that $e_{(f)}$ can be written $e_{(f)} = R_{T}
e_{(i)}$,
where $R_{T}$ is time-reversal); also, it follows that when we propagate some
tetrad
$e_{(i)}^{\prime}$ around ${\gamma}^{\prime}$ the final tetrad is related to
the
initial one by some reflection, $R_{L}$, about a spacelike axis $L$, i.e.,
$e_{(f)}^{\prime} = R_{L} e_{(i)}^{\prime}$. In terms of the pinors ${\psi},
{\psi}^{\prime}$ representing $e, e^{\prime}$ (respectively) we then have
(using now
gamma matrix notation since our pin group is Cliffordian)}
\[
\begin{array}{ccc}
{\psi}_{f} = {\gamma}_{0} {\psi}_{i} & ~~({\mbox{on}} ~{\gamma}) & \\
 & & \\
{\psi}_{f}^{\prime} = {\gamma}_{L} {\psi}_{i}^{\prime} & ~~({\mbox{on}}
{}~{\gamma}^
{\prime}) & \hspace*{8.7cm} (13)
\end{array}
\]
{where of course ${\gamma}_{0}$ represents time reflection and ${\gamma}_{L}$
represents reflection about axis $L$. Now, since we have chosen $a = +, b = -$
recall
that we have}
\[
\begin{array}{cc}
{\gamma}_{0}^{2} = -{\mbox{Identity}} = -I & \hspace*{1cm} ~~~~~{\mbox{and}} \\
 & \\
{\gamma}_{L}^{2} = +{\mbox{Identity}} = I . &
\end{array}
\]

{We now wish to view ${\gamma}$ and ${\gamma}^{\prime}$ as the initial and
final
curves in the two families of curves ${\{}{\gamma}(v)|v ~{\in}~ [0, 1]{\}}$ and
${\{}{\gamma}^{\prime}(v)|v ~{\in}~ [0, 1]{\}}$ which sweep out $T$ and
$T^{\prime}$
respectively. We are then concerned with the following question: To what extent
is an
anomaly on $T$ or $T^{\prime}$ determined simply by insisting that $w_{1}^{-}
{}~{\smile}~ w_{1}^{-} ~[T] ~{\not=}~ 0$ or $w_{1}^{+} ~{\smile}~ w_{1}^{+}
{}~[T^{\prime}]
{}~{\not=}~ 0$ and that $a = +, b = -$ ?

First consider the curves ${\gamma}^{\prime}(v)$ sweeping out $T^{\prime}$.
Suppose
(for the purpose of contradiction) that there was an anomaly. Then we would
have}
\[
\begin{array}{lc}
{\psi}^{\prime}_{f}(0) = {\gamma}_{L} {\psi}_{i}^{\prime}(0) & \hspace*{9.2cm}
(14) \\
 & \\
{\psi}^{\prime}_{f}(1) = -{\gamma}_{L} {\psi}_{i}^{\prime}(1) & \hspace*{9.2cm}
(15)
\end{array}
\]

{Furthermore, because $w_{1}^{+} ~{\smile}~ w_{1}^{+} ~[T^{\prime}] ~{\not=}~
0$ it
follows that there is a curve $c^{\prime}(v)$ in $T^{\prime}$ (generated by the
parameter $v$) with the property that propagating tetrads around $c^{\prime}$
{\it also} reverses spacelike orientation, i.e., we have}
\[
\begin{array}{cc}
{\gamma}_{L} {\psi}^{\prime}_{f}(1) = {\psi}^{\prime}_{f}(0) & \hspace*{9.2cm}
(16) \\
 & \\
{\psi}^{\prime}_{i}(1) = {\gamma}_{L} {\psi}_{i}^{\prime}(0) & \hspace*{9.2cm}
(17)
\end{array}
\]

{However, combining equations (15) and (17) we obtain}
\[
{{\psi}^{\prime}_{f}(1) = -{\gamma}_{L} {\gamma}_{L} {\psi}_{i}^{\prime}(0) =
-{\gamma}_{L} {\psi}_{f}^{\prime}(0)}
\]
{but this contradicts (16) ! Thus, we see that $w_{1}^{+} ~{\smile}~ w_{1}^{+}
{}~[T^{\prime}] ~{\not=}~ 0$ together with ${\gamma}_{L} {\gamma}_{L} = +I$
imply that we
cannot have an anomaly (arising from the `parity reversal' part of some
arbitrary pin
transform). But this is exactly why $w_{1}^{+} ~{\smile}~ w_{1}^{+}$ is not
relevant
to the obstruction class. If $w_{1}^{+} = 0$, then ${\gamma}_{L}$ does not even
arise
in our considerations, and so the question of an anomaly in ${\gamma}_{L}$
becomes
moot.

On the other hand, consider the curves ${\gamma}(v)$ sweeping out $T$.
{\it Now} suppose (for the purpose of contradiction) that there is
{\it no} anomaly. Then we have}
\[
\begin{array}{cc}
{\psi}_{f}(0) = {\gamma}_{0} {\psi}_{i}(0) & \hspace*{9.2cm} (18) \\
 & \\
{\psi}_{f}(1) = {\gamma}_{0} {\psi}_{i}(1) & \hspace*{9.2cm} (19)
\end{array}
\]

{Furthermore, the assumption that $w_{1}^{-} ~{\smile}~ w_{1}^{-} ~[T]
{}~{\not=}~ 0$
again implies}
\[
\begin{array}{cc}
{\gamma}_{0} {\psi}_{f}(1) = {\psi}_{f}(0) & \hspace*{9.2cm} (20) \\
 & \\
{\psi}_{i}(1) = {\gamma}_{0} {\psi}_{i}(0) & \hspace*{9.2cm} (21)
\end{array}
\]

{However, combining equations (19) and (21) now gives us}
\[
{{\psi}_{f}(1) = {\gamma}_{0} {\gamma}_{0} {\psi}_{i}(0) = -{\psi}_{i}(0) =
{\gamma}_{0} {\psi}_{f}(0)}
\]
{However, equation (20) is equivalent to}
\[
{{\psi}_{f}(1) = -{\gamma}_{0} {\psi}_{f}(0)}
\]
{and so again we have a contradiction. But this means that $w_{1}^{-}
{}~{\smile}~
w_{1}^{-} ~[T] ~{\not=}~ 0$ together with ${\gamma}_{0}^{2} = -I$ imply that we
{\it must} have an anomaly on $T$ ! But this is exactly why $w_{1}^{-}
{}~{\smile}~ w_{1}^{-}$ {\it is} relevant to the obstruction class. In other
words, we have shown the following}\\
\\
{\noindent {\it Fact}~~ Suppose there is a two-cycle, $T$, in $M$ such that
$w_{1}^{d} ~{\smile}~ w_{1}^{d} ~[T] ~{\not=}~ 0$ (where $d$ can be $+$ or
$-$). Then
the values of $a, b ~{\in}~ {\{}{\pm}{\}}$ {\it alone} can affect the
anomalous behaviour of pinor fields on $T$ (and hence on $M$).

Indeed, we now see that the above constructions can be used to rederive the
results of
Section III.

Now, however, let us briefly go one step further and analyse the breakdown of
pin
structure by generalising a construction of Witten [14] (see also [15] and [16]
for
related reading).

Recall that Witten (in [14]) interprets the breakdown of spin structure in
terms of
anomalies in the fermion effective action which arises when one quantizes a
point
particle with world line supersymmetry.

Explicitly, Witten takes the world line of the particle to be a closed curve
${\gamma}$ in the spacetime. He then constructs the fermion effective action,
${\sqrt{{\det}(D)}}~({\gamma})$, where $D$ is the `world line Dirac operator',}
\[
{D = i \left(\begin{array}{c}
{\frac{d}{dt}} ~{\delta}^{i}_{j} ~+~ {\frac{dx^{l}}{dt}} ~S^{i}_{lj}
\end{array}\right)}
\]
{where $S_{ilj}$ is the spin connection. Thus, to define ${\sqrt{{\det}(D)}}$
we need
only know the eigenvalues of $D$. Now, let us (following Witten) just consider
the
relationship between anomalies in ${\sqrt{{\det}(D)}}$ and the breakdown of
{\it Pin(4)} structure; that is, we do not decompose the tangent bundle
${\tau}_{M}$ into `spacelike' and `timelike' parts determined by some Lorentz
structure on $M$ (i.e., we are concentrating here simply on lifting the $O(4)$
structure of the tangent bundle). Then the first thing we must recall is that
there
are {\it two} types of Pin(4)-structure, which we write ${\mbox{Pin}}^{+}(4)$
and ${\mbox{Pin}}^{-}(4)$. ${\mbox{Pin}}^{+}(4)$ is the $2 - 1$ cover of $O(4)$
with
the property that the element ${\gamma}_{+}$, which generates the non-identity
connected component of ${\mbox{Pin}}^{+}(4)$, satisfies ${\gamma}^{2}_{+} =
{\mbox{Id}}$. The obstruction to ${\mbox{Pin}}^{+}(4)$ structure on $M$ can
then be
calculated using the above constructions, and we can see that the obstruction
is that
the following hold:}
\[
{w_{2}({\tau}_{M}) = 0}
\]
{(See also [17] for another derivation).

On the other hand, ${\mbox{Pin}}^{-}(4)$ is the $2 - 1$ cover of $O(4)$ with
the
property that the element ${\gamma}_{-}$, which generates the non-identity
connected
component of ${\mbox{Pin}}^{-}(4)$, satisfies ${\gamma}^{2}_{-} =
-{\mbox{Id}}$. Thus,
the obstruction to ${\mbox{Pin}}^{-}(4)$ structure is that the following hold:}
\[
{w_{2}({\tau}_{M}) ~+~ w_{1}({\tau}_{M}) ~{\smile}~ w_{1}({\tau}_{M}) = 0 .}
\]

{Now, as Witten notes, with the choice of $O(4)$ for tangent bundle structure
group we
have that $A^{i}_{j} = {\frac{dx^{l}}{dt}} ~S_{lj}^{i}$ is an $O(4)$-invariant
gauge
field on ${\gamma}$. We are concerned with how choosing our pin group (i.e.,
boundary
conditions) affects this gauge field and hence the eigenvalues of $D$ (and thus
the
value of det$(D)$). To see how this happens, let us again consider an explicit
example.

First, let $M$ be a manifold with $w_{2}({\tau}_{M}) = w_{1}({\tau}_{M})
{}~{\not=}~ 0$
and let there be a 2-cycle $T$ in $M$ such that $w_{1} ~{\smile}~ w_{1} ~[T]
{}~{\not=}~
0$. Then $M$ does admit ${\mbox{Pin}}^{-}(4)$ structure but does not admit
${\mbox{Pin}}^{+}(4)$ structure. This means we must differentiate between the
determinants used in the two situations; thus, let ${\det}^{\pm}(D)$ denote the
determinants obtained using the groups ${\mbox{Pin}}^{\pm}(4)$, respectively.

Now, for ${\det}^{+}(D)$ we see that we get the same boundary condition as the
one that Witten considers (i.e., he takes his gamma matrix to have square equal
to
{\it plus} the identity). It follows that $A$ can be gauge transformed into the
form}
\[
{A = {\frac{1}{2{\pi}}} \left( \begin{array}{cc}
\begin{array}{cc}
0 &{\theta}_{1} \\
-{\theta}_{1} &0
\end{array} &{\mbox{\LARGE 0}} \\
{\mbox{\LARGE 0}} &\begin{array}{cc}
0 &{\theta}_{2} \\
-{\theta}_{2} &0
\end{array}
\end{array} \right)}
\]
{regardless of the fact that $w_{1} ~{\smile}~ w_{1} ~[T] ~{\not=}~ 0$ (here we
are
again regarding $T$ as being swept out by a continuous family of world lines
${\gamma}(v)$). Thus, using [14] we obtain}
\[
{{\sqrt{{\det}^{+}(D)}} = {\prod_{i = 1}^{2}} ~{\sin}\left(\begin{array}{c}
{\frac{{\theta}_{i}}{2}} \end{array}\right)  \hspace*{8cm} (22)}
\]
{The relevance of the fact that $w_{1} ~{\smile}~ w_{1} ~[T] ~{\not=}~ 0$
becomes
clear when we realise that the form of ${\sqrt{{\det}^{-}(D)}}$ is exactly the
same as
expression (22), but the boundary conditions satisfied by the angles
${\theta}_{1},
{\theta}_{2}$ are different. Explicitly, the total amount that the angles
change in
${\sqrt{{\det}^{-}(D)}}$ (as we interpolate from ${\gamma}(0)$ to
${\gamma}(1)$) must
differ from the amount they change in ${\sqrt{{\det}^{+}(D)}}$ by ${\pi}$.

In our example, we are assuming there is an anomaly in ${\sqrt{{\det}^{+}(D)}}$
(i.e.,
there is no ${\mbox{Pin}}^{+}(4)$ structure). It follows that one of the angles
must
change by $2{\pi}$ while the other stays fixed, that is, we must have something
like}
\[
\begin{array}{l}
{\theta}_{1}(0) = {\theta}_{1}(1) \\
{\theta}_{2}(0) = {\theta}_{2}(1) ~+~ 2{\pi}
\end{array}
\]

{However, the angles appearing in ${\sqrt{{\det}^{-}(D)}} ~(1)$ have an extra
${\pi}$
added in. But this means that the total change in {\underline{both}} angles
appearing
in ${\sqrt{{\det}^{-}(D)}}$ is essentially ${\pi}$, and so there is no anomaly
in the
expression ${\prod_{i = 1}^{2}} ~{\sin}\left(\begin{array}{c}
{\frac{{\theta}_{i}}{2}}
\end{array}\right)$. Thus, we see that as expected there is an anomaly in
${\sqrt{{\det}^{+}(D)}}$ but not ${\sqrt{{\det}^{-}(D)}}$.

When $w_{1} ~{\smile}~ w_{1} = 0$ then the boundary conditions are the same for
both
pin structures (which is what we expect since the obstruction classes are
identical
when $w_{1} ~{\smile}~ w_{1} = 0$).}\\

\vspace*{0.6cm}

{\noindent \bf VI. ~Format for solving the general problem}\\

{In the general situation, we will be given a manifold $M$ with tangent bundle
${\tau}_{M}$ an `O'-bundle satisfying ${\pi}_{1}(O) ~{\simeq}~ G ~{\not\simeq}~
{\{}1{\}}$. We are then concerned with globally lifting ${\tau}_{M}$ to another
bundle
with structure group ${\bar O}$ satisfying}
\[
{1 ~{\longrightarrow}~ {\pi}_{1}(O) ~{\longrightarrow}~ {\bar O}
{}~{\longrightarrow}~ O
{}~{\longrightarrow}~ 1 .}
\]

{Using the Theorem from Section II, we will again obtain a commutative diagram
of
sheaf cohomology groups.

If ${\delta}_{2}:~ H^{1}(M; {\cal O}) ~{\longrightarrow}~ H^{2}(M;
{\pi}_{1}(O))$ is
the Bockstein homomorphism for the vertical sequence in the diagram (as above)
and
${\xi} ~{\in}~ H^{1}(M; {\cal O})$ denotes a choice of principal $O$-bundle,
then we
see that the obstruction to lifting to a principal ${\bar O}$-bundle is now the
element ${\delta}_{2}({\xi}) ~{\in}~ H^{2}(M; {\pi}_{1}(O))$.

For example, if we take a four-manifold with `Kleinian' metric $g_{ab}$
(signature $(+
+ - -)$) then ${\tau}_{M}$ has structure group $O(2, 2)$ satisfying
${\pi}_{1}(O(2,
2)) ~{\simeq}~ {\Bbb Z}$  $~{\times}~ {\Bbb Z}$. Thus, the obstruction to
representing
(globally) the information in ${\tau}_{M}$ in a simply connected way is, in
this case,
an element of $H^{2}(M; {\Bbb Z} ~{\times}~ {\Bbb Z})$. The point is, we could
again write
out the general form of this obstruction, and then use the commutative diagram
of
cohomology groups (analogous to (6)) to calculate the explicit form the
obstruction
takes in the various cases corresponding to how the `discrete' part of $O(2,
2)$ is
covered by the discrete part of the cover, ${\bar O}(2, 2)$.

Finally, some readers may be worried about how far we have extended the
vertical
sequences in (6). However, it is shown ([6], pg. 207) that we can always extend
as far
as we need to (i.e., to $H^{2}(M; {\pi}_{1}(O))$) as long as ${\pi}_{1}(O)$ is
abelian
(regardless of whether or not $C^{a, b, c}$ is abelian).}\\
\vspace*{0.6cm}

{\noindent \bf VII. ~Conclusion}\\

{Finally, we mention one further application of these results, namely, the
calculation
of amplitudes in the Hartle-Hawking approach to treating gravitation in a
quantum-mechanical way ([12], [13]).

Recall that in this approach the basic idea is to take Feynman's `sum over
histories'
philosophy to its logical conclusion, in other words, we sum over manifolds as
well as
metrics. We allow the topology of the universe to fluctuate. More explicitly,
suppose
that ${\{}({\Sigma}_{j}^{i}, {\psi}_{j}^{i}, h_{j}^{i})|j = 1, ... n{\}}$ is a
collection of three-manifolds ${\Sigma}_{1}^{i}, {\Sigma}_{2}^{i}, ...
{\Sigma}_{n}^{i}$ with matter fields ${\psi}_{j}^{i}$ and three metrics $h_{j}$
(representing an `initial' configuration) and ${\{}({\Sigma}_{k}^{f},
{\psi}_{k}^{f},
h_{k}^{f})|k = 1, ... m{\}}$ is a collection of three-manifolds
${\Sigma}_{1}^{f},
{\Sigma}_{2}^{f}, ... {\Sigma}_{m}^{f}$ with matter fields ${\psi}_{k}^{f}$ and
three
metrics $h_{k}$ (representing a `final' configuration). Then the amplitude to
go from
the initial state to the final state is given, in this picture, by}
\[
{<({\Sigma}^{f}, {\psi}^{f}, h^{f})|({\Sigma}^{i}, {\psi}^{i}, h^{i})> ~=
{\displaystyle{\sum_{M}}~{\nu}(M){\int_{\cal C}}{\delta}{\phi}{\delta}g{\rm
e}^{(-I{\{}{\phi}, g, M{\}})}}}
\]
{where the sum is over all manifolds $M$ with boundary}
\[
{{\partial}M ~{\cong}~ {\Sigma}_{1}^{f} ~{\cup}~ {\Sigma}_{2}^{f} ~{\cup}~ ...
{}~{\cup}~ {\Sigma}_{m}^{f} ~{\cup}~ {\Sigma}_{1}^{i} ~{\cup}~ {\Sigma}_{2}^{i}
{}~{\cup}~
... ~{\cup}~ {\Sigma}_{n}^{i},}
\]
{weighted by ${\nu}(M)$, with $I$ the Euclidean action for matter fields
${\phi}$ and
metrics $g$ on $M$ inducing the given configurations on the boundary. The point
is, we
might want to use the `selection rules' derived above (Section IV) to assign
`weight
zero' (${\nu}(M) = 0$) to those manifolds which do not admit some pin or spin
structure, i.e., which are not ${\mbox{Pin}}^{a, b, c}(p, q)$ (or spin)
cobordisms for
the boundary three-manifolds. If we demand the three surfaces ${\Sigma}^{f},
{\Sigma}^{i}$ be everywhere spacelike, then ${\mbox{kink}}({\partial}M; g_{ab})
= 0$
and so we see that such restrictions would be perhaps non-trivial. The precise
effect
such a procedure would have on the class of manifolds appearing in the path
integral
is, however, at present unclear.}\\
\vspace*{0.6cm}

{\noindent \bf Acknowledgements}\\

{The author would like to thank his supervisor, Dr. G.W. Gibbons, for helpful
comments
and guidance during the preparation of this work. Also thanks to Dr. L.
Dabrowski for
useful conversations. Finally, gratitude goes to Jo Ashbourn (Piglit) for
loving
support and help with the preparation of this paper.  This work was supported
by
NSF Graduate Fellowship No. RCD-9255644.}\\
\vspace*{0.6cm}

{\noindent \bf References}\\

{\noindent [1] Chamblin H.A.: Some Applications of Differential Topology in
General Relativity. DAMTP preprint.}\\
\\
{\noindent [2] Dabrowski L.: Group Actions on Spinors. Monographs and Textbooks
in Physical Science, Bibliopolis, 1988}\\
\\
{\noindent [3] Karoubi M.: Algebres de Clifford et K-Theory. Ann. Scient. Ec.
Norm. Sup. $1^{\rm e}$ serre, t.1, pg. 161, 1968}\\
\\
{\noindent [4] Michel L.: Invariance in Quantum Mechanics and Group Extension,
from Group Theoretical Concepts and Methods in Elementary Particle Physics, ed.
Gursen
et al., Gordon {\&} Breach, 1969}\\
\\
{\noindent [5] Wells, Jr., R.O.: Differential Analysis on Complex Manifolds.
2nd
ed., Springer-Verlag, Berlin, New York, 1980}\\
\\
{\noindent [6] Ward R.S., Wells, Jr., R.O.: Twistor Geometry and Field Theory.
Cambridge Monographs in Mathematical Physics, CUP, 1990}\\
\\
{\noindent [7] Geroch R.: Spinor Structure of Spacetimes in General Relativity.
J. Math. Phys. {\bf 9}, 1739--1744, 1968}\\
\\
{\noindent [8] Milnor J.W., Stasheff J.: Characteristic Classes. Princeton
Univ. Press, Princeton, New Jersey, 1974}\\
\\
{\noindent [9] Gibbons G.W., Hawking S.W.: Selection Rules for Topology
Change. Commun. Math. Phys., {\bf 148}, NO. 2, 1992}\\
\\
{\noindent [10] Arnold V.I.: Singularity Theory. London Math. Soc. Lecture
Notes, No. {\bf 53}, CUP, 1981}\\
\\
{\noindent [11] Kervaire M.A., Milnor J.W.: Groups of Homotopy Spheres: I.
Annals of Math,. {\bf 77}, No. 3, 1963}\\
\\
{\noindent [12] Gibbons G.W., Hartle J.B.: Real tunneling geometries and the
large-scale topology of the universe. Phys. Rev. D, {\bf 42}, No. 8, pg. 2458,
1990}\\
\\
{\noindent [13] Hawking S.W.: in 300 Years of Gravitation. ed. S.W. Hawking and
W. Israel, CUP, 1987}\\
\\
{\noindent [14] Witten E.: Global Anomalies in String Theory. Argonne-Chicago
Symposium on Geometry, Anomalies and Topology, 1985}\\
\\
{\noindent [15] Witten E.: An SU(2) Anomaly. Phys. Lett. {\bf 117} B, pg. 324,
1982}\\
\\
{\noindent [16] Hawking S.W., Pope C.N.: Generalised spin structures in quantum
gravity. Phys. Lett. {\bf 73} B, pg. 42, 1978}\\
\\
{\noindent [17] Kirby R.C., Taylor L.R.: Pin Structures on Low-Dimensional
Manifolds. London Math. Soc. Lecture Notes, No. {\bf 151}, CUP, 1989}\\

\end{document}